# Rocking of rigid blocks: Analytical solution and exact energy-based overturning criteria

**Aristotelis E. Charalampakis[1,3], George C. Tsiatas[2], and Panos Tsopelas[3]**

[1]Department of Civil Engineering, University of West Attica, Athens, GR 12244, Greece, https://orcid.org/0000-0001-9072-6259

[2]Department of Mathematics, University of Patras, Rio, GR 26504, Greece, https://orcid.org/0000-0003-4808-7881

[3]School of Applied Mathematical and Physical Sciences, National Technical University of Athens, Athens, GR 15773, Greece, https://orcid.org/0000-0002-1864-0001

## Abstract

In this paper, the underlying energy flow in rocking dynamics of rigid blocks is utilized to elucidate their response details and overturning conditions. Based merely on the law of the conservation of energy, the exact criteria for a rigid block to overturn in the free vibration regime are established for the first time in the context of both nonlinear and linear theory. That is, if the initial conditions of the free oscillation regime fall within the prescribed stable libration region, then it is *a priori* determined that the block will not overturn and *vice versa*. The analytical solution to the nonlinear free rocking problem is also derived, utilizing integration techniques originating from the solution of the nonlinear pendulum. Additionally, pulse-type overturning spectra are constructed easily by numerical integration of the equation of motion only during the pulse duration. If the block does not topple during the forced regime, the angle and angular velocity at the end of the excitations are employed to assess immediately the safety or unsafety of the block using the exact overturning nonlinear criterion. Finally, new insights into the stability regions of the overturning spectra are illuminated and explained in detail using a color mapping.

## 1. Introduction

Since 1963, the year that Housner published his seminal paper on "*The behavior of inverted pendulum structures during earthquakes*" [1], many scholars have been trying to unravel the mystery of stability of free-standing rigid blocks under transient excitation. The ample seismic stability of tall slender structures has been confirmed in a series of experimental and numerical studies, e.g. [2-21]. For further reading, we refer the interested readers to consult the excellent paper by Makris [22], not only for an extensive literature review but also for a thorough investigation of rocking isolation as an alternative mechanism for large slender structures.

However, revisiting the literature on rocking stability it is interesting to notice that there is not a sound criterion - even for the linearized equation of motion - whether the block will eventually topple during the free vibration regime, merely from its initial conditions. It is quite impressive that this plausible question has not been answered yet for one of the oldest man-made earthquake-resistant mechanisms. Housner in 1963, within the context of the linear theory, was the first who derived analytical expressions for the minimum base acceleration for various types of horizontal ground motions required to overturn the rigid block from rest (zero initial conditions) [1]. First, for a rectangular pulse, he correctly postulated that the condition for overturning is that the total work done by the inertial force is just equal to the difference in potential energy between positions of the angle of rotation $\theta = 0$ and $\theta = a$ with $a = \arctan(b/h)$ being the block slenderness (Figure 1). However, the solution to the equation of motion he derived was unfortunately incorrect, which led to an unrealistic overturning condition for the rectangular pulse [equation (9) in his paper]. Secondly, constructing the overturning spectra for a half-sine pulse, Housner falsely postulated that the condition for overturning is that the angle of rotation $\theta$ is equal to $a$ at the time that the half-sine expires [22,23].

The overturning condition presented for the first time by Shi *et al.* [23] [equation (13) in their paper] is actually correct (as will be shown later in Section 2) only for the case that the block overturns without impact in the free vibration regime. Although, the authors correctly stated that "*The minimum overturning condition requires that $\dot{\theta} = 0$ at $\theta = a$; that is, the rocking block at the time of overturning must have nearly zero angular velocity*", it is quite bizarre, however, that this condition has been presented without any explanation and/or discussion of its origin. This was addressed in a sequel work of Anooshehpoor *et al.* [24] who mentioned that this condition originated from the total conservation energy. Applying the overturning condition, Shi *et al.* [23] determined the minimum overturning acceleration amplitude $a_p$ of a half-sine pulse by solving a transcendental equation numerically. The same equation was derived independently by Makris & Roussos [15,16], who stated that a rigid block, with frequency parameter $p$, "*is excited by the minimum overturning acceleration pulse the time needed to reach the verge of overturning $\theta(t_{ov}) = a$ is theoretically infinite; and therefore $tan(pt_{ov}) = 1$*" [22]. The two overturning conditions provided by Shi *et al.* [23] and Makris & Roussos [15,16] are equivalent and valid only under the assumption that no energy is lost during impacts.

An analytical study, in the realm of the linearized equations of motion, was presented by Voyagaki *et al.* [25] who revisited the rocking problem of a rigid, free-standing block subjected to a range of idealized single-lobe ground acceleration pulses. They stated that "*The exact overturning criterion can be developed using the notion of simultaneous immobility $\dot{\theta}(\tau_c) = 0$ and equilibrium $\ddot{\theta}(\tau_c) = 0$ (or $\theta(\tau_c) = a$) of the block, at a critical time $\tau = \tau_c$*", which is essentially the same provided by Shi *et al.* [23] and Makris & Roussos [15,16]. In this regard, they produced overturning spectra for exponential, rectangular, and triangular pulses. Furthermore, Dimitrakopoulos & DeJong [26] and Dimitrakopoulos & Fung [27] derived closed-form solutions that completely describe the overturning of the linearized rocking block under simple pulse excitations and a family of multi-lobe pulse ground motions, respectively. This was achieved by solving the associated transcendental equations to calculate the time of impact, either exactly or approximately, and identifying all the transient overturning modes utilizing the criterion of simultaneous immobility and equilibrium [17,25].

This work provides the missing stepping stone in understanding rocking dynamics and overturning rigid blocks by utilizing the underlying energy flow. The paper is structured as follows. In Section 2, the exact criteria for a rigid block to overturn in the free vibration regime are established for the first time based solely on the principle of conservation of energy in the context of both nonlinear and linear theory. Stemming from the oscillatory study of nonlinear pendulums, the energy boundary which separates the phase space into two distinct regions is termed the separatrix energy $E_{sep}$. In the eye-shaped region (Figure 2a) inside the separatrix, the pendulum stably librates (oscillates back and forth), while in the region outside the separatrix the pendulum swings over (traversing unstable paths). In the rocking motion of rigid blocks without loss of energy during impacts, the separatrix has a rhomboidal shape (Figure 2b) that corresponds to stable libration inside it, while the region outside it represents the motion's toppling paths. On the other hand, the energy loss during consecutive impacts will diminish the amplitude of oscillation since a certain level of energy loss occurs at every impact of the rigid block. It will be shown that only the energy loss on the first impact is crucial for the stability of the rigid block, leading to certain stable (= non-toppling) oscillations for angles well beyond the angular limit $\pm a$. This fact results in an extension of the libration zone, forming two antisymmetric blade-like regions which complement the initial rhomboidal region. If and only if the initial conditions of the free oscillation fall within the prescribed libration region, then it is *a priori* determined that the block will not overturn. In Section 3 the analytical solution to the nonlinear free rocking problem is derived, utilizing techniques from the solution of the nonlinear pendulum [28–30]. The solution is given in segments, separated by the time instants when $\theta = 0$, since energy loss occurs at each impact. Section 4 is devoted to the construction of the overturning spectra for the one-sine acceleration pulse (type-A=forward pulse), the one-cosine pulse (type-B=forward-and-back pulse), and the symmetric and antisymmetric Ricker wavelets. If the block does not topple during the forced vibration regime, then $\theta_0$ and $\dot{\theta}_0$ are evaluated at the end of the pulse and the numerical integration is terminated. At this point, the safety or unsafety of the block is assessed immediately, without any further numerical integration, using the exact criterion presented in Section 2. Using color mapping, the stability regions of the overturning spectra are identified and explained in detail. Finally, certain conclusions drawn from this study are summarized in Section 5.

## 2. Problem statement –Energy conservation principle

Consider the free undamped rocking motion of a solid homogeneous rectangular block with dimensions $2b \times 2h$, as shown in Figure 1, where $R = \sqrt{b^2 + h^2}$ is the half-diagonal of the rectangle (or the distance of the centroid of mass from the pivot point $O$), and $a = \arctan(b/h)$ is the block slenderness. Since sliding is prevented, the motion is described solely by the angle $\theta$ and its initial conditions $\theta_0$, $\dot{\theta}_0$ at time instant $t_0 = 0$. For negative values of $\theta$, rotation takes place around the pivot point $O'$.

The equation that governs the free oscillation can be derived using the energy conservation principle. The total mechanical energy of the system is the sum of its kinetic energy:

$$E_k = \frac{1}{2} I_O \dot{\theta}^2, \tag{1}$$

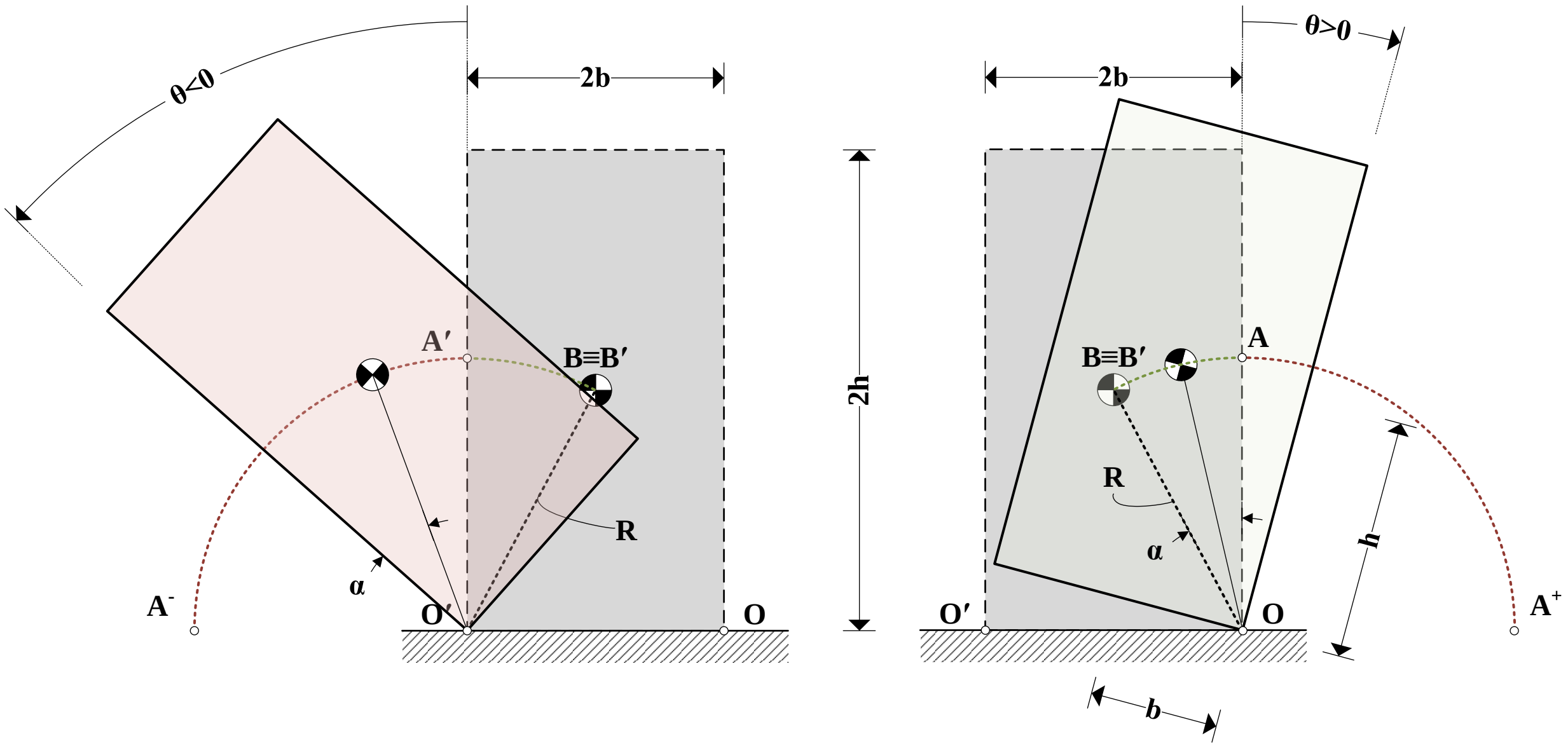

**Figure 1** Free oscillation of a solid homogeneous rectangular block.

where, $\dot{\theta}$ is the angular velocity and $I_O = \frac{4}{3} mR^2$ is the mass moment of inertia of the rigid block about each pivot point, and its potential energy at an angle $\theta$:

$$E_p = R[\cos(\alpha - |\theta|) - \cos a]mg, \tag{2}$$

which is taken 0 when $\theta = 0$ i.e., the potential energy datum is at $\theta = 0$ when the rigid block is upright. In the absence of non-conservative forces, the total mechanical energy is conserved at every time instant including the energy at $t_0$:

$$E_k + E_p = E_{k,0} + E_{p,0} = \text{const.} \tag{3}$$

Utilizing the initial conditions in Eqs. (1), (2) and substituting into Eq. (3) leads to:

$$\dot{\theta}^2 = \dot{\theta}_0^2 + \frac{3g}{R}\left[\sin^2\left(\frac{\alpha-|\theta|}{2}\right) - \sin^2\left(\frac{\alpha-|\theta_0|}{2}\right)\right], \tag{4}$$

which after the introduction of the *frequency parameter* $p = \sqrt{3g/4R}$ becomes:

$$\dot{\theta}^2 = \dot{\theta}_0^2 + 4p^2\left[\sin^2\left(\frac{\alpha-|\theta|}{2}\right) - \sin^2\left(\frac{\alpha-|\theta_0|}{2}\right)\right]. \tag{5}$$

Eq. (5) is the governing energy conservation equation, holding at all times, and can be used to generate the orbits of rocking rigid blocks in the phase space. From Eq. (5), the Euler-Lagrange well-known free vibration equation of rocking motion of a rigid block is obtained [1]:

$$\ddot{\theta}(t) = -p^2 \sin[a \operatorname{sign}\theta(t) - \theta(t)]. \tag{6}$$

An interesting inherent trait of pendulum-type motions is the energy boundary that separates the two modes of behavior. For a pendulum with oscillation limits between $-\pi$ and $\pi$ (see Figure 2a) this energy is termed separatrix energy $E_{sep}$ and separates the phase space into two distinct regions, i.e., the light blue region inside the separatrix where the pendulum librates (i.e., oscillates back and forth) and the region outside the separatrix where the pendulum swings over.

## 2.1 Rocking motion without energy loss during impacts

First, we discuss the rocking response of rigid blocks under the assumption that no energy is lost during impacts (occurring whenever $\theta = 0$ for $t > 0$). In this case, the separatrix is the curve passing through points $A$ (and $A'$) of maximum potential energy and zero angular velocity, and points $B$ (and $B'$) of maximum kinetic energy and zero potential energy (see Figure 1, Figure 2b). Using expression (5) we can now plot the transition among the points $A \rightarrow B' \rightarrow A' \rightarrow B \rightarrow A$ which leads to the rhomboidal separatrix of Figure 2b, where the light blue region corresponds to the stable libration rocking motion and the region outside the separatrix represents the motion's toppling paths. Note that the phase space portrait of the rocking motion (Figure 2b) can be created from the phase space of the pendulum (Figure 2a) by putting together the two orange strips having a width equal to $a$. The separatrix energy, which is constant for every point on rhomboidal separatrix of Figure 2b, can be determined using Eq. (2) for point $A$ (or $A'$) where toppling is imminent with $\theta = a$, and $\dot{\theta} = 0$ as:

$$E_{sep} = R(1 - \cos a)mg. \tag{7}$$

To identify the maximum initial angular velocity that can be applied to an upright resting block ($\theta_0 = 0$), so that it will not overturn, the conservation of energy between points $A$ ($\theta_0 = a$, $\dot{\theta}_0 = 0$) and $B'$ ($\theta = 0$, $\dot{\theta} = -\dot{\theta}_{\max}$), i.e., the application of Eq. (5), gives:

$$\dot{\theta}_{\max} = 2p\sin\left(\frac{\alpha}{2}\right), \tag{8}$$

which for slender blocks (i.e., small angles $a$) can be linearized to $\dot{\theta}_{\max} = pa$ [6,31], considering $\sin\left(\frac{\alpha}{2}\right) \approx \frac{\alpha}{2}$.

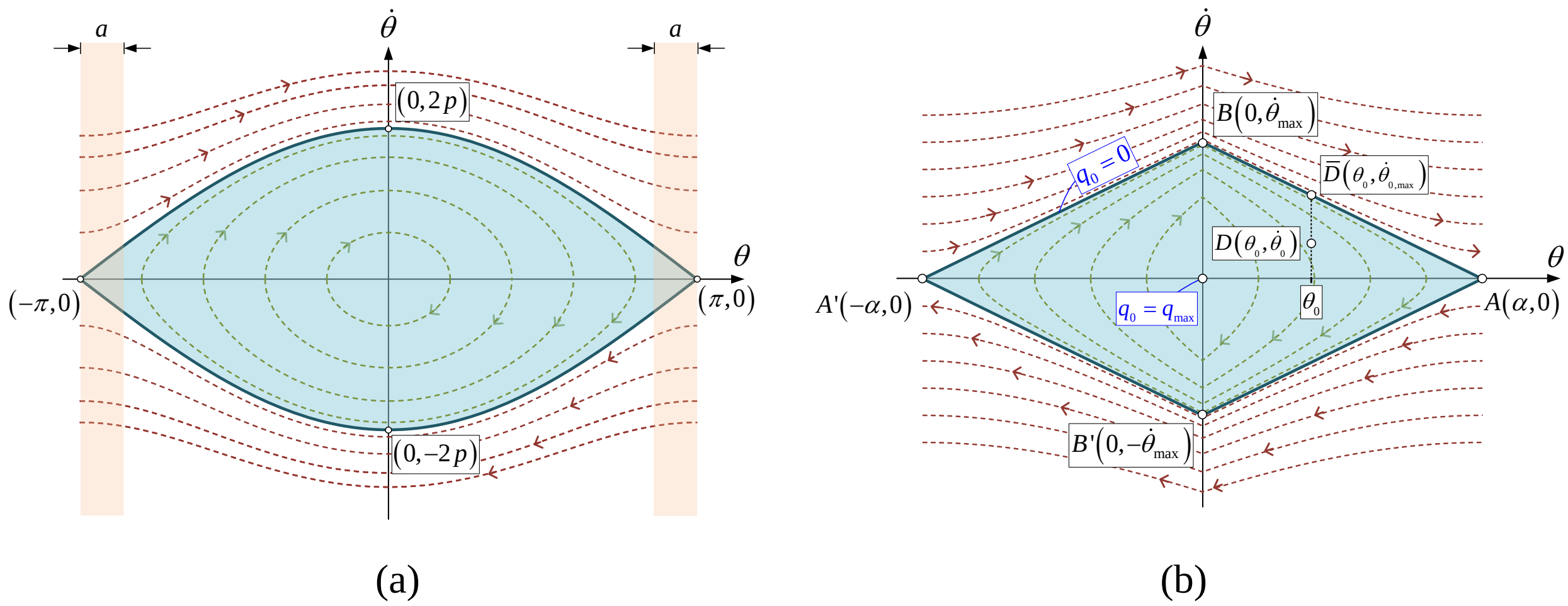

**Figure 2** Phase space portrait of (a) pendulum with oscillation limits between $-\pi$ and $\pi$, and (b) rocking rigid block with no energy loss during impacts.

As already mentioned, the rigid block's initial energy $E_0$ has as upper bound for stable rocking motion which is the separatrix energy $E_{sep}$. This requirement, using Eqs. (1) through (3), takes the form:

$$R(1-\cos a)mg \geq \frac{1}{2}I_O\dot{\theta}_0^2 + R[\cos(a-|\theta_0|) - \cos a]mg. \tag{9}$$

At the edge of overturning Eq. (9) can be written as:

$$mgR[1-\cos(a-|\theta_0|)] = \frac{1}{2}I_O\dot{\theta}_0^2, \tag{10}$$

which is in fact the overturning condition presented for the first time by Shi *et al.* [23] [equation (13) in their paper]. Although, the authors correctly stated that "*The minimum overturning condition requires that* $\dot{\theta} = 0$ *at* $\theta = a$*; that is, the rocking block at the time of overturning must have nearly zero angular velocity*", Eq. (10) is valid only for the case when the block overturns without impact in the free vibration regime.

From Eq. (9) we can derive the exact criterion for the stable free rocking motion of a free rigid block set in motion by its initial conditions only, under the assumption that no energy is lost during impacts:

$$4p^2\sin^2\left(\frac{a-|\theta_0|}{2}\right) - \dot{\theta}_0^2 \geq 0 \quad \text{or} \quad 2p\sin\left(\frac{a-|\theta_0|}{2}\right) - \left|\dot{\theta}_0\right| \geq 0. \tag{11}$$

Setting now

$$q_0 = 4p^2\sin^2\left(\frac{a-|\theta_0|}{2}\right) - \dot{\theta}_0^2, \tag{12}$$

the criterion to ensure stability, i.e., that the *rigid block stably librates* (without toppling), reads:

$$q_0 \geq 0, \tag{13}$$

while when $q_0 < 0$ the rectangular block topples. Note that the term $\frac{1}{2} I_O q_0$ has units of energy; thus, the parameter $q_0$ can be considered as a measure of the difference between the system's initial state energy and the separatrix energy, and, consequently, it appears multiple times in the analytical solution of the nonlinear free rocking problem presented in the next section. For engineering applications, the exact overturning criterion Eq. (11) can be approximated with:

$$p(a - |\theta_0|) - |\dot{\theta}_0| \geq 0. \tag{14}$$

It should be noted that the two overturning criteria [viz. Eqs. (11) and (14)] are practically identical for slenderness angle $a < 40^o$. In this case, the general rhomboidal shape of the separatrix (Figure 2b) degenerates to an actual rhomb with vertices $A, B', A', B$.

Using the limiting case of $q_0 = 0$, (i.e., points on the separatrix) one can evaluate the maximum value of $\dot{\theta}_{0,max}$ that can be applied to the block so that it will not topple when $\theta_0 \neq 0$ (see points $D$ and $\overline{D}$ in Figure 2b):

$$\dot{\theta}_{0,max} = 2p \sin\left(\frac{a - |\theta_0|}{2}\right). \tag{15}$$

The maximum value of $q_0$, obtained for an upright block at rest (i.e., at the origin of the graph of Figure 2b), is:

$$q_{\max} = 4p^2 \sin^2\left(\frac{a}{2}\right) = \dot{\theta}_{\max}^2. \tag{16}$$

## 2.2 Rocking motion with energy loss during impacts

In this subsection, the assumption that no energy is lost during impacts (imposed in the previous section) is lifted. It will be shown that the energy loss on the first impact leads to certain stable (= non-toppling) oscillations for angles well beyond the angular limit $\pm a$.

The energy loss due to impacts depends on many factors, as discussed in [32–34]. Even the presence of a foreign object beneath the rocking block, or defects in the interface between the ground and the block, will alter the position of the impulsive forces experienced by the block during impact. Naturally, this will greatly affect the energy loss during impact, and even a non-symmetric behavior may arise for the two directions of motion. A concave bottom surface of the block will force the position of the impulsive forces at the pivot points, following the assumption by Housner (1963). Alternatively, the use of devices and recesses at the pivot points will ensure contact strictly at the pivot points, no sliding, and sufficient energy loss during impacts.

During the rocking motion, a certain level of energy loss occurs at every impact of the rigid block. The angular velocity after each impact is reduced by $\sqrt{r}$. The parameter $r$ is the established coefficient of restitution, which is determined by the conservation of angular momentum just before and right after the impact [1]:

$$r = \left(1 - \frac{3}{2}\sin^2 a\right)^2. \tag{17}$$

This is the maximum value under which a block with slenderness $a$ will undergo rocking motion [6,10]. If additional energy is lost during the impact, then the true coefficient of restitution will be even less. Potential sliding, or rocking and sliding, during impact, can be prevented with a recess at the pivot points [22].

As expected, the energy loss during consecutive impacts will diminish the amplitude of oscillation. In the question of how this affects the stability of the rigid block, the answer is that *only* the first impact is crucial. As shown in Figure 3, a rigid block at point $\bar{B}$ (or $\bar{B}'$) just before the first impact has the maximum (or minimum) angular velocity that a block can attain without overturning in the rest of the motion. In other words, if the angular velocity is larger than $\dot{\theta}_{\max}/\sqrt{r}$ (or smaller than $-\dot{\theta}_{\max}/\sqrt{r}$) just before the first impact, then the block will overturn. This fact results in an extension of the libration zone, forming two antisymmetric blade-like regions which complement the initial rhomboidal region. If the initial conditions of the free oscillation fall within the libration region, then it is *a priori* guaranteed that the block will not overturn (see Figure 3). Each "blade" of the extended libration region is narrow when $\theta_0$ is $\pm\pi/2$, but it is quite wide around values $\pm a$. Thus, slender columns are actually more stable than indicated by their geometry and there are initial angular velocities that can drive the rigid block to safe libration motion even when the initial angle of rotation is $\theta_0 = \pm\pi/2$.

Thus, taking into account the energy loss during the first impact and if $\theta_0\dot{\theta}_0 < 0$, i.e., if the initial angle of rotation and angular velocity have opposite signs, then the minimum value of $q_0$ for non-toppling behavior is not zero but rather given by:

$$q_{\min} = 4p^2 \sin^2\left(\frac{a}{2}\right)\left(1 - \frac{1}{r}\right) = q_{\max}\left(1 - \frac{1}{r}\right) < 0. \tag{18}$$

All points identified in Figure 3 are evaluated using energy conservation, i.e., application of Eq. (5). For example, the conservation of energy between points $B$ ($\theta = 0$, $\dot{\theta} = \dot{\theta}_{\max}/\sqrt{r}$) and $\bar{A}'$ ($\theta = -a$) yields:

$$\dot{\theta}_{\bar{A}'} = \dot{\theta}_{\max}\sqrt{\frac{1}{r} - 1}. \tag{19}$$

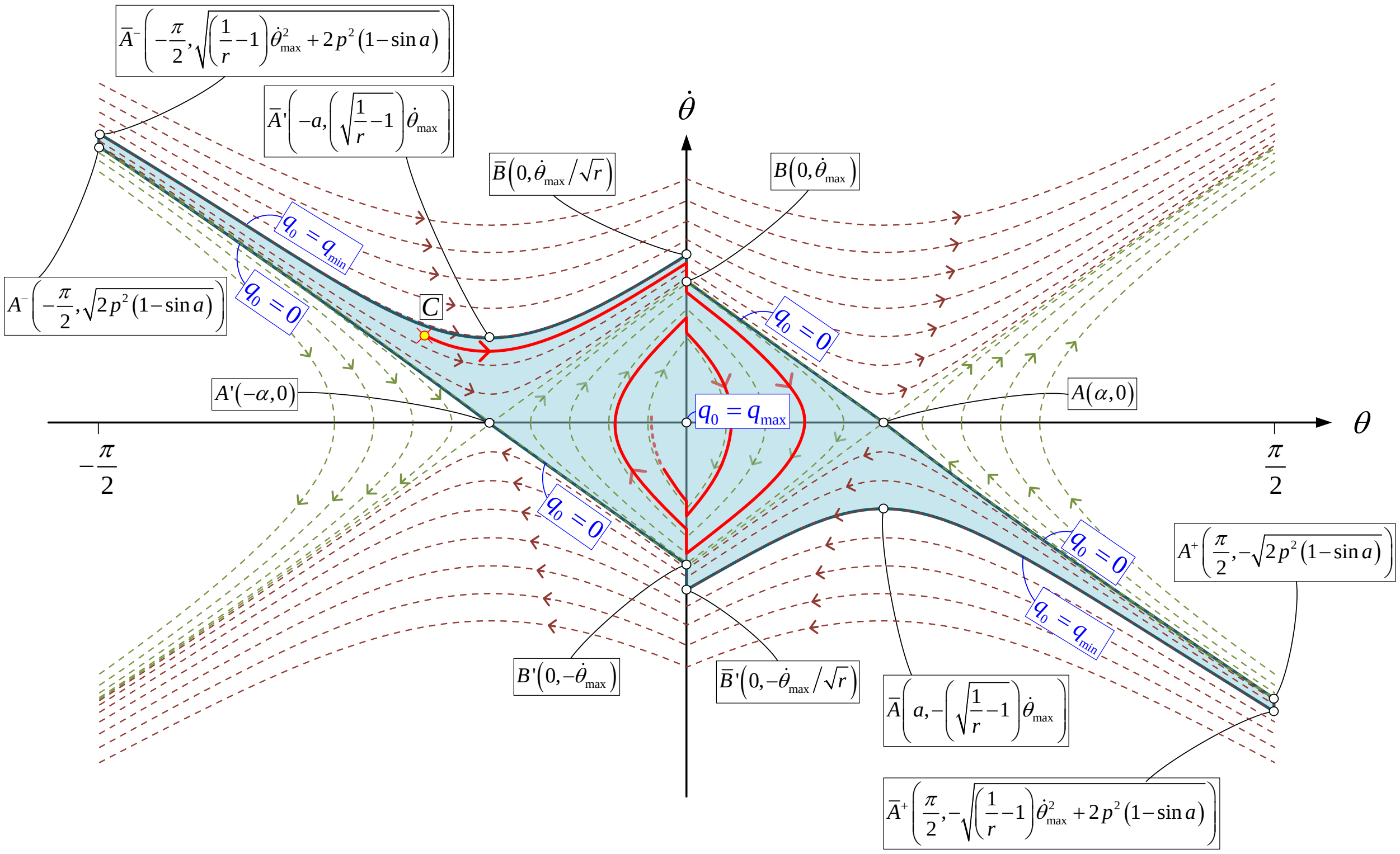

**Figure 3** Phase space portrait of rocking rigid block with energy loss during impacts. The rhomboidal safe libration region of Figure 2b is extended due to energy loss on the first impact. The light blue libration region is expressed by the exact stability criterion (20).

Observing Figures 3 and 2b, it is clear that the blade-like regions, essentially formed by the reduction of angular velocity in the transition of $\bar{B} \rightarrow B$ (and $\bar{B}' \rightarrow B'$) during the first impact, include stable rocking orbits that previously led to the toppling of the rigid block. For instance, an initial point $C$, although well beyond the angular limits $\pm a$, actually leads to the stable rocking of diminishing amplitude (following the red line in Figure 3). Under the assumption that no energy is lost during impacts, the same point $C$ will lead to the overturning of the rigid block.

Eventually, the phase space can be divided into seven distinct regions, as shown in Figure 4. The regions are denoted as safe ("S") or unsafe ("U"), depending on whether the rigid block will topple or not. Thus, the three safe regions are: ***S***, corresponding to the rhomboidal stable libration region, ***S+***, and ***S-*** corresponding to the blade-like parts of the extended libration region, with positive or negative angular velocities, respectively. The four unsafe regions are termed ***U0+***, ***U1+***, ***U0-***, ***U1-***. They contain either a "0" or "1" to signify whether they overturn with no impact or a single impact, respectively. Note that overturn after multiple impacts during free oscillation is not possible. The "+" or "**-**" signify the sign of the angular velocity, and the side to which the block will eventually topple over.

The final exact stability condition of the block is given by the following expression:

$$\left(-\frac{\pi}{2} \le \theta_0 \le \frac{\pi}{2} \wedge \theta_0\dot{\theta}_0 < 0 \ \wedge q_{\min} \le q_0 \le 0\right) \vee (-a \le \theta_0 \le a \ \wedge 0 \le q_0 \le q_{\max}) \ \Leftrightarrow \ \text{Stability}, \quad (20)$$

where "∨" signifies the logical "OR" and "∧" the logical "AND". It should be noted that the first parenthesis of the criterion (20) refers to the blade-like ***S+*** and ***S-*** regions while the second one refers to the rhomboidal ***S*** region.

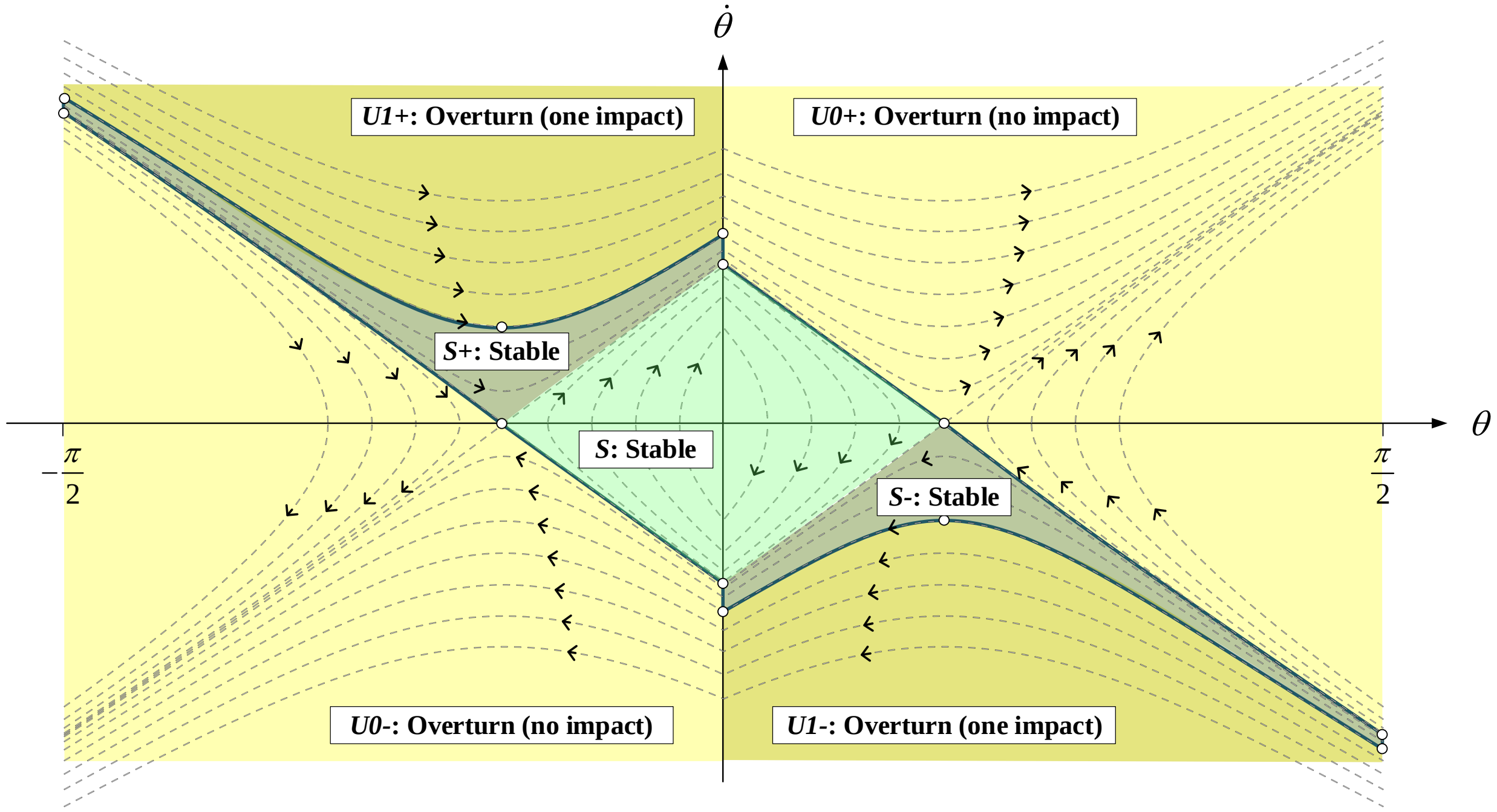

**Figure 4** Division of the phase space into seven regions, depending on the outcome of the free rocking motion. Green and dark green regions are safe; yellow and dark yellow regions are unsafe.

## 3. Analytical solution to the nonlinear free rocking problem

Eq. (5), after the substitution of expression (12), takes the form:

$$\dot{\theta}^2 = 4p^2 \sin^2\left(\frac{\alpha - |\theta|}{2}\right) - q_0. \quad (21)$$

The nonlinear ordinary differential Eq. (21) is integrated analytically for $\theta > 0 \wedge \theta_0 > 0$ utilizing techniques used in the derivation of the solution for the nonlinear pendulum [28–30]. Defining the new variable:

$$y = \sin\left(\frac{\alpha - \theta}{2}\right), \quad (22)$$

and differentiating with respect to $t$ reads:

$$\dot{y} = -\frac{1}{2}\cos\left(\frac{\alpha - \theta}{2}\right)\dot{\theta}. \quad (23)$$

Solving the previous relation for $\dot{\theta}$:

$$\dot{\theta} = -\frac{2\dot{y}}{\cos\left(\frac{a-\theta}{2}\right)}, \tag{24}$$

and substituting it into Eq. (21), we have:

$$\frac{4\dot{y}^2}{1-y^2} = 4p^2y^2 - q_0, \tag{25}$$

which after some algebra leads to:

$$\dot{y}^2 = \frac{1}{4}(4p^2y^2 - q_0)(1 - y^2), \tag{26}$$

or:

$$\dot{y} = \frac{dy}{dt} = \pm p\sqrt{\left(y^2 - \frac{q_0}{4p^2}\right)\,(1 - y^2)}. \tag{27}$$

Separating the variables and integrating both parts yields:

$$\int_{y_0}^{y} \frac{dy}{\sqrt{\left(y^2-\frac{q_0}{4p^2}\right)(1-y^2)}} = \pm p \int_0^t dt. \tag{28}$$

The integral in the LHS of Eq. (28) involves the use of special functions:

$$\frac{2ip}{\sqrt{q_0}}\left[\mathrm{F}\left(\sin^{-1} y\,, \frac{2p}{\sqrt{q_0}}\right) - \mathrm{F}\left(\sin^{-1} y_0\,, \frac{2p}{\sqrt{q_0}}\right)\right] = \pm pt, \tag{29}$$

where, $u = \mathrm{F}(\varphi, k) = \int_0^{\varphi} \frac{d\theta}{\sqrt{1-k^2\sin^2\theta}}$ with $(-\frac{\pi}{2} < \varphi < \frac{\pi}{2})$ is the incomplete elliptic integral of the first kind. Substituting back $y_0 = \sin\left(\frac{a-\theta_0}{2}\right)$ and $y = \sin\left(\frac{a-\theta}{2}\right)$ yields:

$$\mathrm{F}\left(\frac{a-\theta}{2}, \frac{2p}{\sqrt{q_0}}\right) - \mathrm{F}\left(\frac{a-\theta_0}{2}, \frac{2p}{\sqrt{q_0}}\right) = \pm\frac{\sqrt{q_0}}{2i}t. \tag{30}$$

Solving Eq. (30) for $\theta$ provides the final expression for the angle which is given as:

$$\theta(t) = a + 2\,\mathrm{am}\left[\mp\frac{\sqrt{-q_0}}{2}t + \mathrm{F}\left(\frac{\theta_0-a}{2}, \frac{2p}{\sqrt{q_0}}\right), \frac{2p}{\sqrt{q_0}}\right], \tag{31}$$

where the upper sign corresponds to $\dot{\theta}_0 \geq 0$, the lower sign corresponds to $\dot{\theta}_0 < 0$ and $\varphi = \mathrm{F}^{-1}(u, k) = \mathrm{am}(u, k)$ with $(0 < k^2 < 1)$ is the Jacobi amplitude function with elliptic modulus $k$. The corresponding expression for the angular velocity takes the form:

$$\dot{\theta}(t) = \mp\sqrt{-q_0}\,\mathrm{dn}\left[\mp\frac{\sqrt{-q_0}t}{2} + \mathrm{F}\left(\frac{\theta_0-a}{2}, \frac{2p}{\sqrt{q_0}}\right), \frac{2p}{\sqrt{q_0}}\right], \tag{32}$$

where, $\mathrm{dn}(z, k) = \sqrt{1 - k^2\sin^2[\mathrm{am}(z, k)]}$ with $k^2 < 1$ is the Jacobi DN function.

In what follows, unified expressions of the angle $\theta(t)$ and the angular velocity $\dot{\theta}(t)$ are given which yield the correct result depending on the sign of $\theta_0$ and $\dot{\theta}_0$. The solution is given in segments, separated by the time instants when $\theta = 0$ (see Figure 5 for the nomenclature). This is not a restriction but rather a necessity since during impact a certain level of energy loss occurs. After each impact, the amplitude of angular velocity is multiplied by $\sqrt{r}$, where $r$ is the coefficient of restitution [1]. It is important to clarify that the proposed approach utilizes a given constant coefficient of restitution solely for evaluating purposes. The expressions are easily modified to accommodate a non-constant or even asymmetric coefficient of restitution.

### 3.1 Solution until the first impact

To account for all possible initial conditions, the following auxiliary signum functions are introduced:

$$\mathrm{sgn}_1 = \mathrm{sgn}_1(\theta_0, \dot{\theta}_0) = \mathrm{sign}\left[\mathrm{sign}(\theta_0) + \frac{\mathrm{sign}(\dot{\theta}_0)}{2}\right], \tag{33}$$

$$\mathrm{sgn}_2 = \mathrm{sgn}_2(\theta_0, \dot{\theta}_0) = \mathrm{sign}\left[\mathrm{sign}(\theta_0)\,\mathrm{sign}(\dot{\theta}_0) + \frac{1}{2}\right], \tag{34}$$

where:

$$\mathrm{sign}(x) = \begin{cases} +1, & x > 0 \\ 0, & x = 0 \\ -1, & x < 0 \end{cases}. \tag{35}$$

The following expressions are derived, which return the angle $\theta(t)$ and the angular velocity $\dot{\theta}(t)$ for all cases of initial conditions for a non-toppling block, from time $t_0 = 0$ until time $t_1$ of the first impact ($q_{\min} < q_0 < q_{\max} \wedge q_0 \neq 0$):

$$\theta_{01}(t) = \mathrm{sgn}_1\left\{a - 2\,\mathrm{am}\left[\frac{\mathrm{sign}(q_0)\,\mathrm{sgn}_2}{2}\sqrt{-q_0}\,t - \mathrm{F}\left(\frac{|\theta_0| - a}{2}, \frac{2p}{\sqrt{q_0}}\right), \frac{2p}{\sqrt{q_0}}\right]\right\}, \tag{36}$$

$$\dot{\theta}_{01}(t) = -\,\mathrm{sign}(q_0)\,\mathrm{sgn}_1\mathrm{sgn}_2\sqrt{-q_0}\,\mathrm{dn}\left[\mathrm{F}\left(\frac{|\theta_0| - a}{2}, \frac{2p}{\sqrt{q_0}}\right) - \frac{\mathrm{sign}(q_0)\,\mathrm{sgn}_2}{2}\sqrt{-q_0}\,t, \frac{2p}{\sqrt{q_0}}\right], \tag{37}$$

The time instant of the first impact $t_1$ is given by:

$$t_1 = -\frac{2\,\mathrm{sign}(q_0)}{\sqrt{-q_0}}\left[\mathrm{F}\left(\frac{a}{2}, \frac{2p}{\sqrt{q_0}}\right) - \mathrm{sgn}_2\,\mathrm{F}\left(\frac{|\theta_0| - a}{2}, \frac{2p}{\sqrt{q_0}}\right) - 2\,\mathrm{H}(q_0)\,\mathrm{dn}^{-1}\left(0, \frac{2p}{\sqrt{q_0}}\right)\right], \tag{38}$$

where, $\mathrm{H}(x)$ is the Heaviside (step) function. The angular velocity $\dot{\theta}_{1p}$ *prior* to the first impact can be evaluated by utilizing Eq. (5):

$$\dot{\theta}_{1p} = -\mathrm{sgn}_1\sqrt{4p^2 \sin^2\left(\frac{\alpha}{2}\right) - q_0}. \tag{39}$$

## 3.2 Solution after the first impact

The rest of the oscillation can be evaluated as follows, taking into account the energy loss during impacts. Let $n = 1,2, ....$ be the number of impacts. The angular velocity immediately after the $n^{th}$ impact is given by:

$$\dot{\theta}_{na} = (-1)^{n-1}\, r^{n/2}\dot{\theta}_{1p}, \tag{40}$$

where, without loss of generality, we assume that the coefficient of restitution is constant. The value of parameter $q$ after the $n^{th}$ impact:

$$q_n = 4p^2 \sin^2\left(\frac{a}{2}\right) - \dot{\theta}_{na}^2. \tag{41}$$

The half-period of oscillation, or equivalently the time between $n^{th}$ and $(n+1)^{th}$ impact:

$$T_{hn} = \frac{4\left[dn^{-1}\left(0, \frac{2p}{\sqrt{q_n}}\right) - \mathrm{F}\left(\frac{a}{2}, \frac{2p}{\sqrt{q_n}}\right)\right]}{\sqrt{-q_n}}. \tag{42}$$

The time of the next impact is found recursively:

$$t_{n+1} = t_n + T_{hn}. \tag{43}$$

When the energy of the system gradually tends to zero because of the consecutive impacts, Eq. (43) converges to an upper limit of $t$, after which all movement practically stops (i.e., $T_{hn} \to 0$). Thus, the time instant of the end of the rocking motion is given as a limit. Finally, the angle $\theta(t)$ and the angular velocity $\dot{\theta}(t)$ between $n^{th}$ and $(n+1)^{th}$ impact are given by:

$$\theta_n(t) = \mathrm{sign}(\dot{\theta}_{na})\left\{\alpha - 2\,\mathrm{am}\left[\frac{\sqrt{-q_n}}{2}(t - t_n) + \mathrm{F}\left(\frac{a}{2}, \frac{2p}{\sqrt{q_n}}\right), \frac{2p}{\sqrt{q_n}}\right]\right\}, \tag{44}$$

$$\dot{\theta}_n(t) = -\,\mathrm{sign}(\dot{\theta}_{na})\sqrt{-q_n}\;\mathrm{dn}\left[\frac{\sqrt{-q_n}}{2}(t - t_n) + \mathrm{F}\left(\frac{a}{2}, \frac{2p}{\sqrt{q_n}}\right), \frac{2p}{\sqrt{q_n}}\right]. \tag{45}$$

The maximum angular acceleration occurs during impacts and it is evaluated from Eq. (6) as:

$$\ddot{\theta}_{\max} = p^2 \sin a. \tag{46}$$

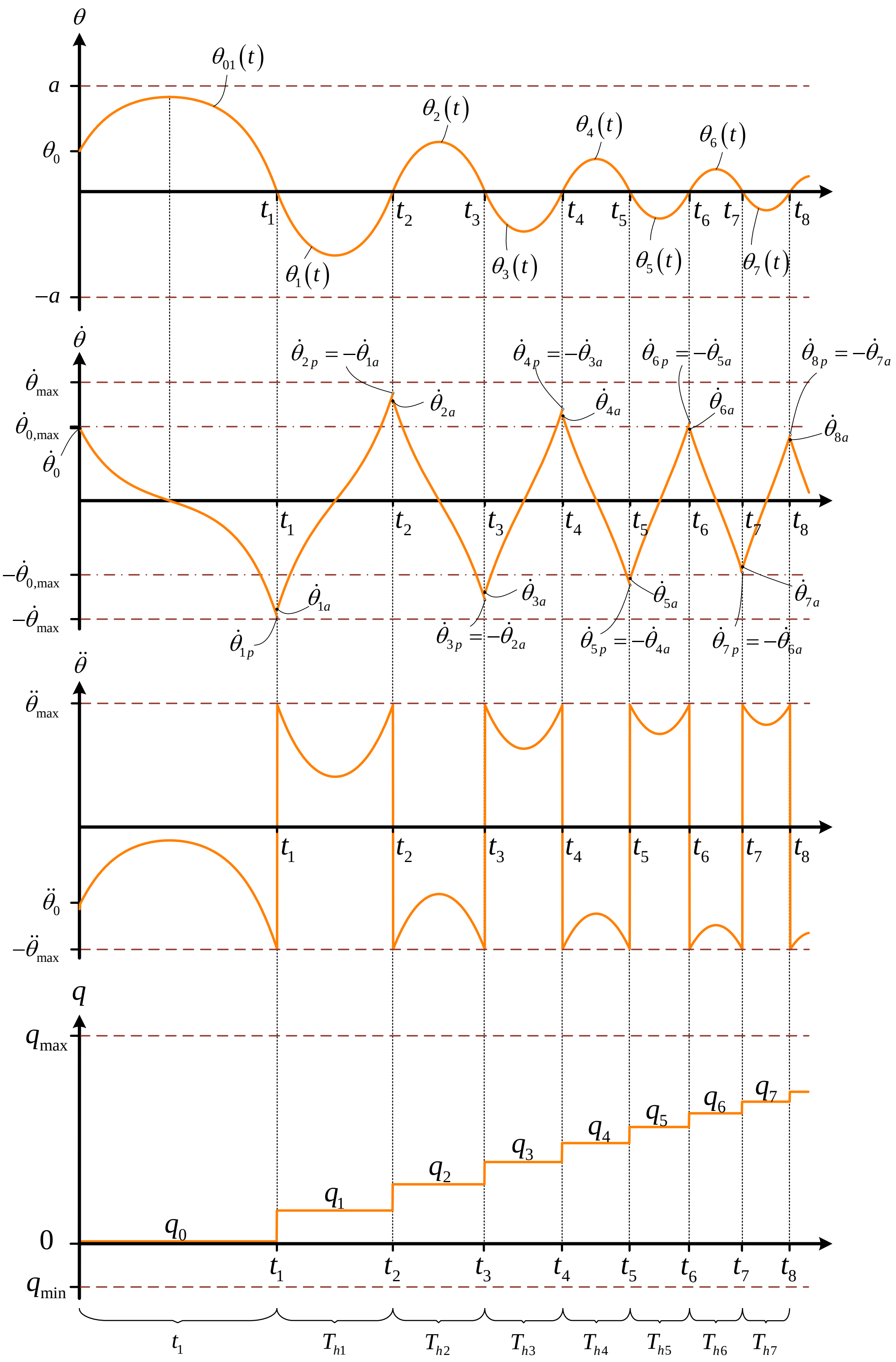

**Figure 5** Nomenclature for the time history of the rotation $\theta(t)$, the angular velocity $\dot{\theta}(t)$, the angular acceleration $\ddot{\theta}(t)$ and the variable $q$.

### 3.3 Implementation details

The implementation of the $\mathrm{am}(u,k)$, $\mathrm{F}(\varphi,k)$, $\mathrm{dn}(u,k)$ and $\mathrm{dn}^{-1}(z,k)$ functions in Mathematica (2019) are $\mathrm{JacobiAmplitude}[u,m]$, $\mathrm{EllipticF}[\varphi,m]$, $\mathrm{JacobiDN}[u,m]$, and $\mathrm{InverseJacobiDN}[\mathrm{z},\mathrm{m}]$ respectively, i.e., the parameter $m=k^2$ is used instead of the elliptic modulus.

In Maple (2020) the respective functions are $\mathrm{JacobiAM}(u,k)$, $\mathrm{InverseJacobiAM}(\varphi,k)$, $\mathrm{JacobiDN}(z,k)$ and $\mathrm{InverseJacobiDN}(z,k)$, as the $\mathrm{EllipticF}(\varphi,k)$ function in Maple actually gives the Legendre normal form of the incomplete elliptic integral of the first kind and not the normal trigonometric form. Also, the $\mathrm{signum}(x)$ function must be used for Eqs. (33) and (34) after setting the variable $_Envsignum0 := 0$.

In all cases, due to numerical precision issues, numbers with infinitesimal imaginary parts may arise where a real number is expected. This can be remedied by using the real part of the complex number.

## 4. Overturning pulse-type spectra

The study of pulses in structural dynamics is quite important since it can give insight into how our structures respond to seismic motions. The overturning criteria presented in Section 2 highly facilitate the construction of the overturning spectra. All that is needed is the evaluation of the angle $\theta_0$ and angular velocity $\dot{\theta}_0$ at the end of the pulse excitations (which coincides with the beginning of the free rocking motion, examined in the previous sections). The forced rocking motion equation:

$$\ddot{\theta}(t) = -p^2\left\{\sin[a\,\mathrm{sign}\,\theta(t)-\theta(t)] + \frac{\ddot{u}_g(t)}{g}\cos[a\,\mathrm{sign}\,\theta(t)-\theta(t)]\right\}, \tag{47}$$

for a given ground motion acceleration $\ddot{u}_g(t)$, is integrated numerically with zero initial conditions in the time interval $0 \le t \le T_f$ with $T_f \le T_P$ being the duration of the excitation after the rocking motion has been initiated. In this work, Eq. (47) has been integrated numerically with the standard ordinary differential equations (ODE) solvers Mathematica (2019) and the results were cross-checked with the available solvers in MATLAB (2016).

If the block does not topple during the forced vibration regime, namely if $-\frac{\pi}{2} < \theta(t) < \frac{\pi}{2} \forall t \in [0, T_f]$, then $\theta_0$ and $\dot{\theta}_0$ are evaluated at the end of the pulse and the numerical integration is terminated. At this point, the safety or unsafety of the block is assessed immediately without any further numerical integration in the free vibration regime. First, the quantities $q_0$, $q_{\min}$, and $q_{\max}$ are evaluated from Eqs. (12), (18), and (16), respectively. The stability (no overturning) of the block is ensured if and only if the following criterion is met: $\left(-\frac{\pi}{2} \le \theta_0 \le \frac{\pi}{2} \wedge \theta_0\dot{\theta}_0 < 0 \wedge q_{\min} \le q_0 \le 0\right) \vee (-a \le \theta_0 \le a \wedge 0 \le q_0 \le q_{\max})$.

In this section, the overturning spectra of pulse-type motions, which have been in the literature to be good representatives of near-source ground motions [e.g. 15,17,37,38], are constructed and analyzed. The rectangular block under consideration in this section has the following geometric properties [22]: $b = 1\,m$, $h = 5.67\,m$, $R = \sqrt{b^2 + h^2} \cong 5.758\,m$, $p = \sqrt{\frac{3g}{4R}} \cong 1.13\,\frac{rad}{s}$, $a = \arctan\left(\frac{b}{h}\right) \cong 0.175\,rad \cong 10\,deg$, $r_{\max} = \left(1 - \frac{3}{2}\sin^2 a\right)^2 \cong 0.912$.

## 4.1 Type-A pulse (one-sine)

The acceleration, velocity, and displacement histories of a one-sine (type-A) are given by Makris & Roussos [16]. The ground acceleration has been shifted so that the instant $t = 0$ corresponds to the initiation of the forced rocking (Figure 6a):

$$\ddot{u}_g(t) = a_p \sin(\omega_p t + \psi), \tag{48}$$

where, $a_p$ is the amplitude, $\omega_p$ is the circular frequency of the sinusoidal pulse, $\psi = \arcsin\left(\frac{g\tan\alpha}{a_p}\right) < \frac{\pi}{2}$ is the phase when the rocking motion is initiated. This is necessary to be taken into account for pulses in which the acceleration builds up gradually, as the one-sine pulse [22]. Obviously, if $a_p < g\tan\alpha$ then rocking is never initiated. The period of the pulse is given as:

$$T_p = \frac{2\pi}{\omega_p}, \tag{49}$$

while the duration of the forced rocking motion is:

$$T_f = \frac{2\pi - \psi}{\omega_p}. \tag{50}$$

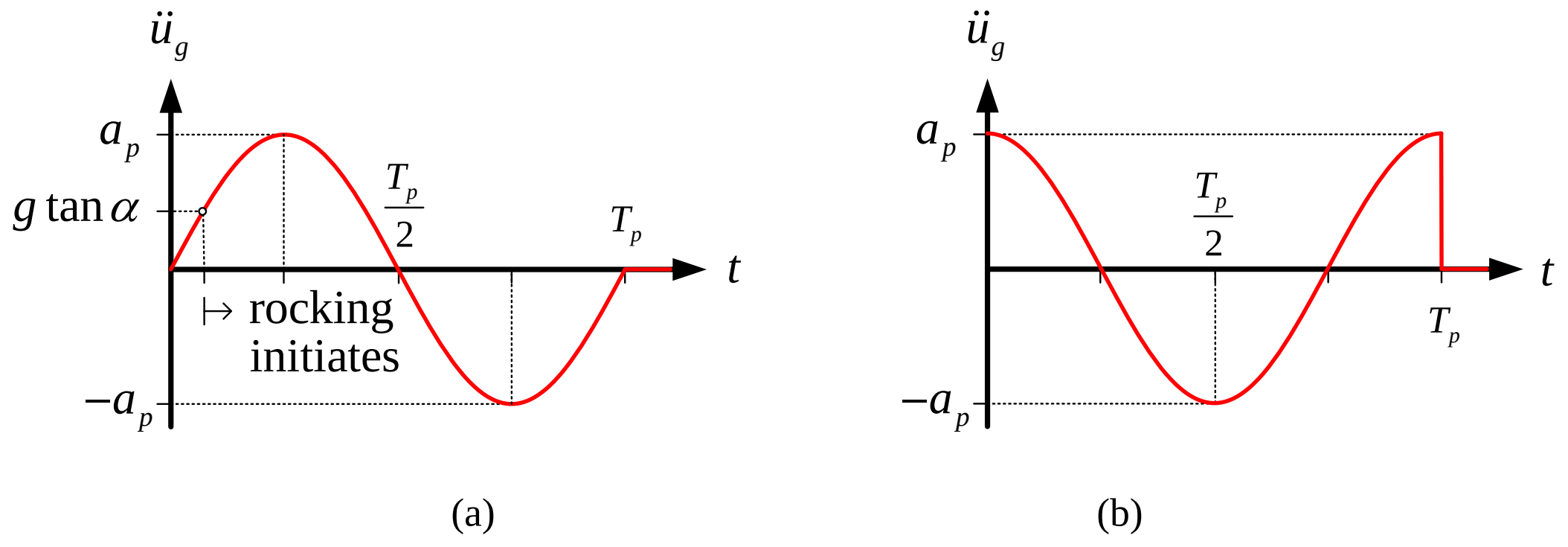

**Figure 6** Acceleration time history of (a) Type-A pulse (one-sine) and (b) Type-B pulse (one-cosine).

Figure 7 shows the overturning spectrum in a 100 × 100 pixels graph. The colors appear in Figure 7a, represent the stable and unstable regions of Figure 4. More specifically, both green (corresponding to the rhomboidal ***S*** region) and dark green (corresponding to the blade-like ***S+*** and ***S-*** regions) are safe regions. Light yellow regions correspond to overturn during free oscillation without impact (***U0+*** and ***U0-*** regions); dark yellow regions correspond to overturn during free oscillation with a single impact (***U1+*** and ***U1-*** regions), while the grey region is reserved for failure during the forced rocking stage. The color code used in Figure 7a reveals that it is actually the blade-like ***S+*** and ***S-*** regions, encoded as dark green, and not the highly nonlinear nature of the response that is responsible for the narrow safe strip dividing the unsafe regions [6]. The same color-coding is used throughout this work concerning the overturning spectra.

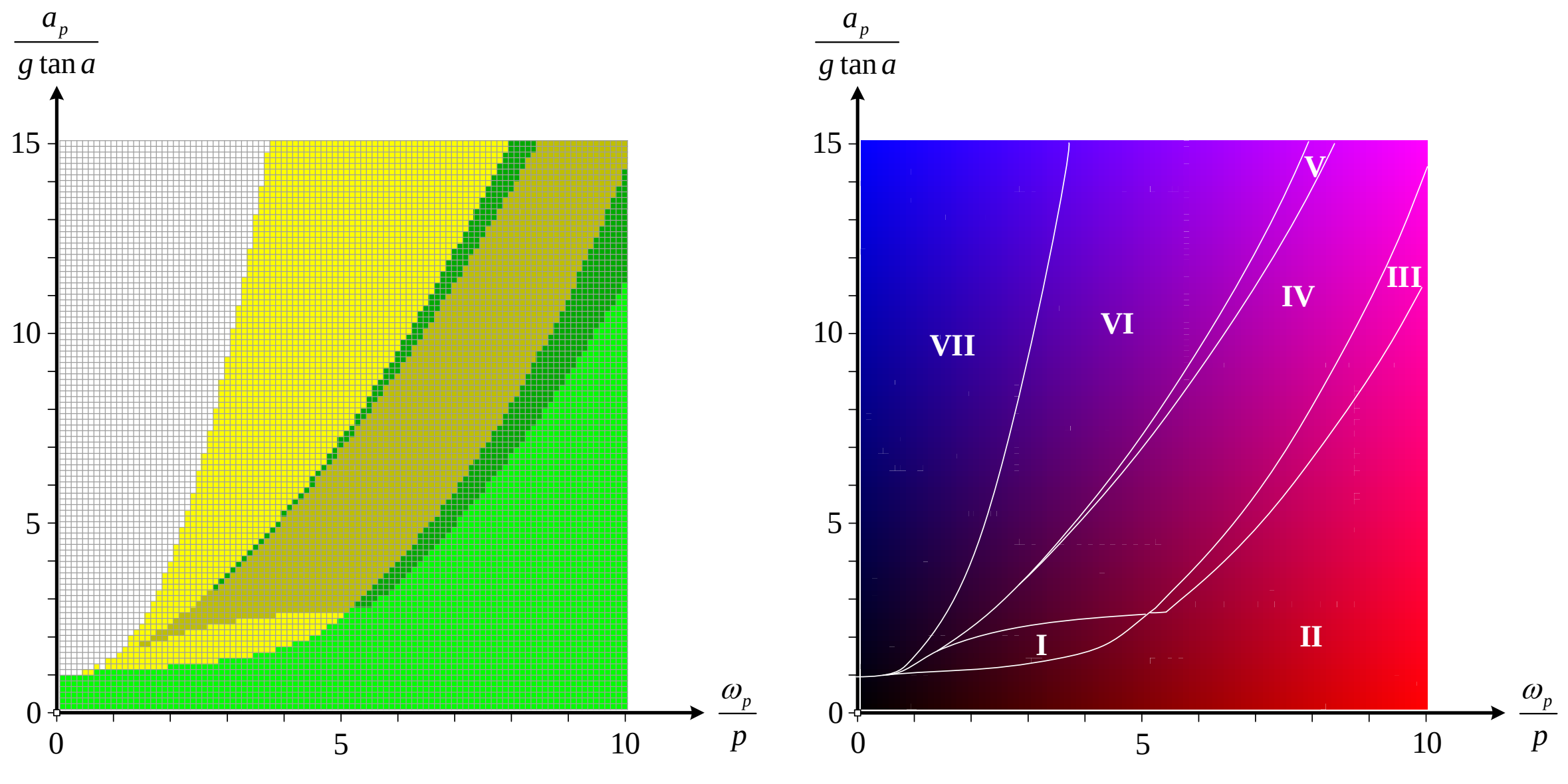

**Figure 7** (a) Overturning acceleration spectrum of a free-standing rectangular block (b) Color mapping of the input $\left(\frac{\omega_p}{p}, \frac{a_p}{g \tan a}\right)$ of the forced stage superimposed over the boundaries of the various regions identified in the overturning spectrum; rectangular block with $a \cong 10\ deg$, $p \cong 1.13\ rad/s$, $r = r_{\max} \cong 0.912$, one-sine pulse.

Further insight on this example can be provided by using "color mapping" to reveal the connection between the input $\left(\frac{\omega_p}{p}, \frac{a_p}{g \tan a}\right)$ and the output $(\theta_0, \dot{\theta}_0)$ of the forced stage of oscillation in phase space. In Figure 7b, the input space has been color mapped as follows: starting with black at the bottom left, red is added along the $x$ axis and blue is added along the $y$ axis. Thus, the bottom right corner is pure red, the top left corner is pure blue, and each point has a unique blue-red color blend. The same color map is used for all cases examined in the rest of the manuscript. The various regions identified in the overturning spectrum of Figure 7a have also been plotted and numbered.

The output $(\theta_0, \dot{\theta}_0)$ of the forced vibration stage shown in Figure 8 reveals the landing spot of each input point $\left(\frac{\omega_p}{p}, \frac{a_p}{g \tan a}\right)$, superimposed over the extended safe libration boundaries, allowing for proper identification of each region. Region I refers to weak to medium-period pulses, landing outside the libration region with $\theta_0 > 0 \wedge \dot{\theta}_0 > 0$, and thus is a ***U0+*** region. Region II mainly refers to medium to short-period pulses, as well as all the pulses that rocking is never initiated ($\frac{a_p}{g \tan a} < 1$). All these cases land inside the rhomboidal ***S*** region. Interestingly, regions III, IV, and V all refer to strong short-period pulses (or large blocks). It is the peculiar shape of the "blade" of the libration region that splits them into these three distinct groups, with regions III and V being ***S+*** (safe) and region IV being ***U1+*** (unsafe). Region VI fails without impact in the free rocking stage and therefore is a ***U0-*** region. Region VII is not shown at all since it fails without ever reaching the free rocking stage (it is located in the extension of region VI in the third quadrant).

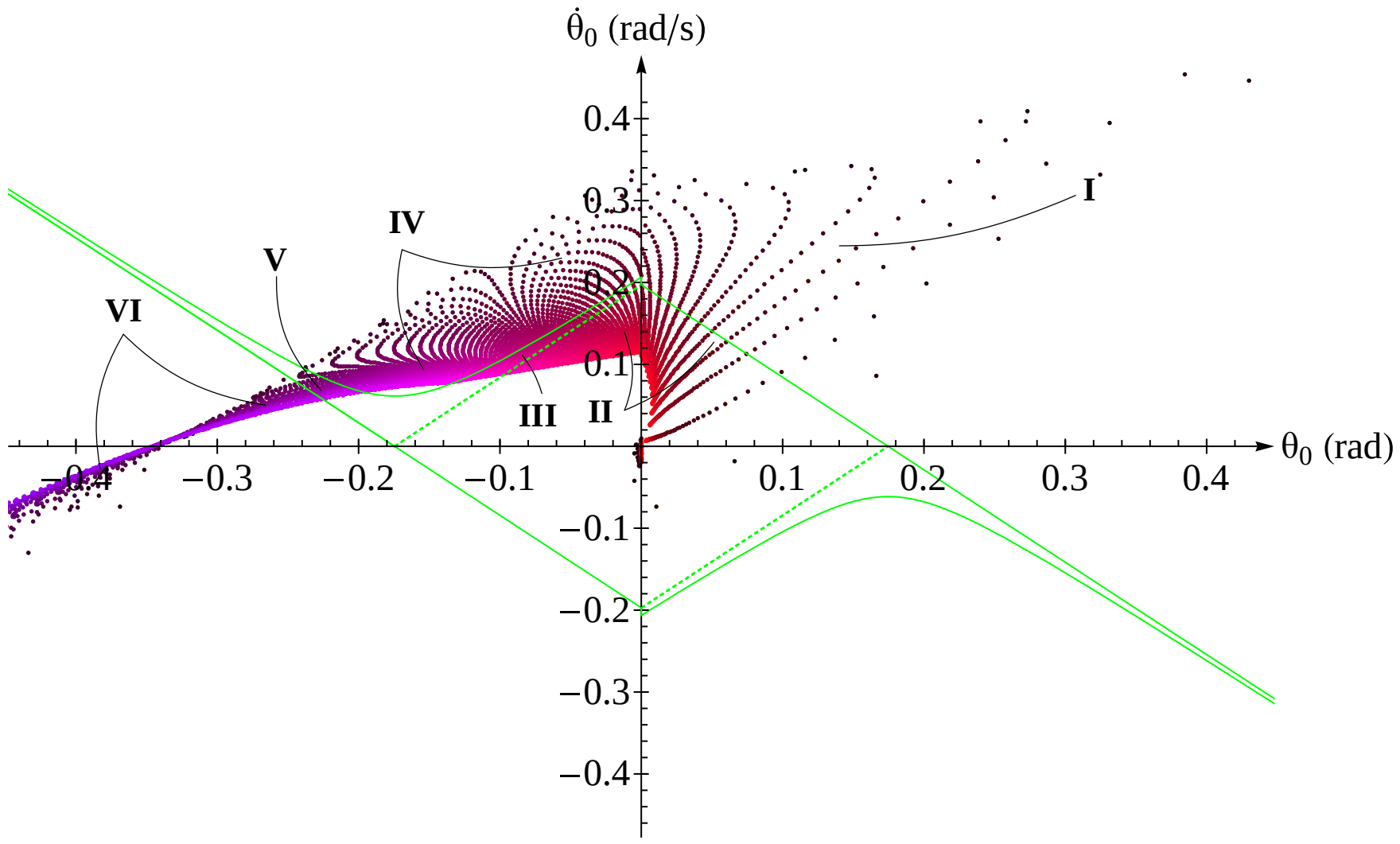

**Figure 8** Output $(\theta_0, \dot{\theta}_0)$ of the forced rocking stage for the overturning spectrum of Figure 7 (one-sine pulse) superimposed over the safe libration boundaries (rectangular block with $a \cong 10\ deg$, $p \cong 1.13\ rad/s$, $r = r_{\max} \cong 0.912$, one-sine pulse).

In all calculations so far, we have utilized the maximum value $r_{\max}$ of the coefficient of restitution [viz. Eq. (17)]. A smaller value of $r$ will not alter the output of the forced vibration stage considerably, because most of the points (practically all the points with $\theta_0 < 0$) will not experience an impact during the forced rocking stage. However, a smaller $r$ will result in thicker ***S+*** and ***S-*** regions, allowing regions III and V to "connect" at the expense of region IV (Figure 8). This is exactly what is observed when constructing the overturn response spectra using 85% and 70% of $r_{\max}$ (Figure 9a and Figure 9b, respectively). Also, note that region I is also diminished because more and more overturning trajectories are now redirected into the safe rhomboidal ***S*** region. This is apparent in Figure 10.

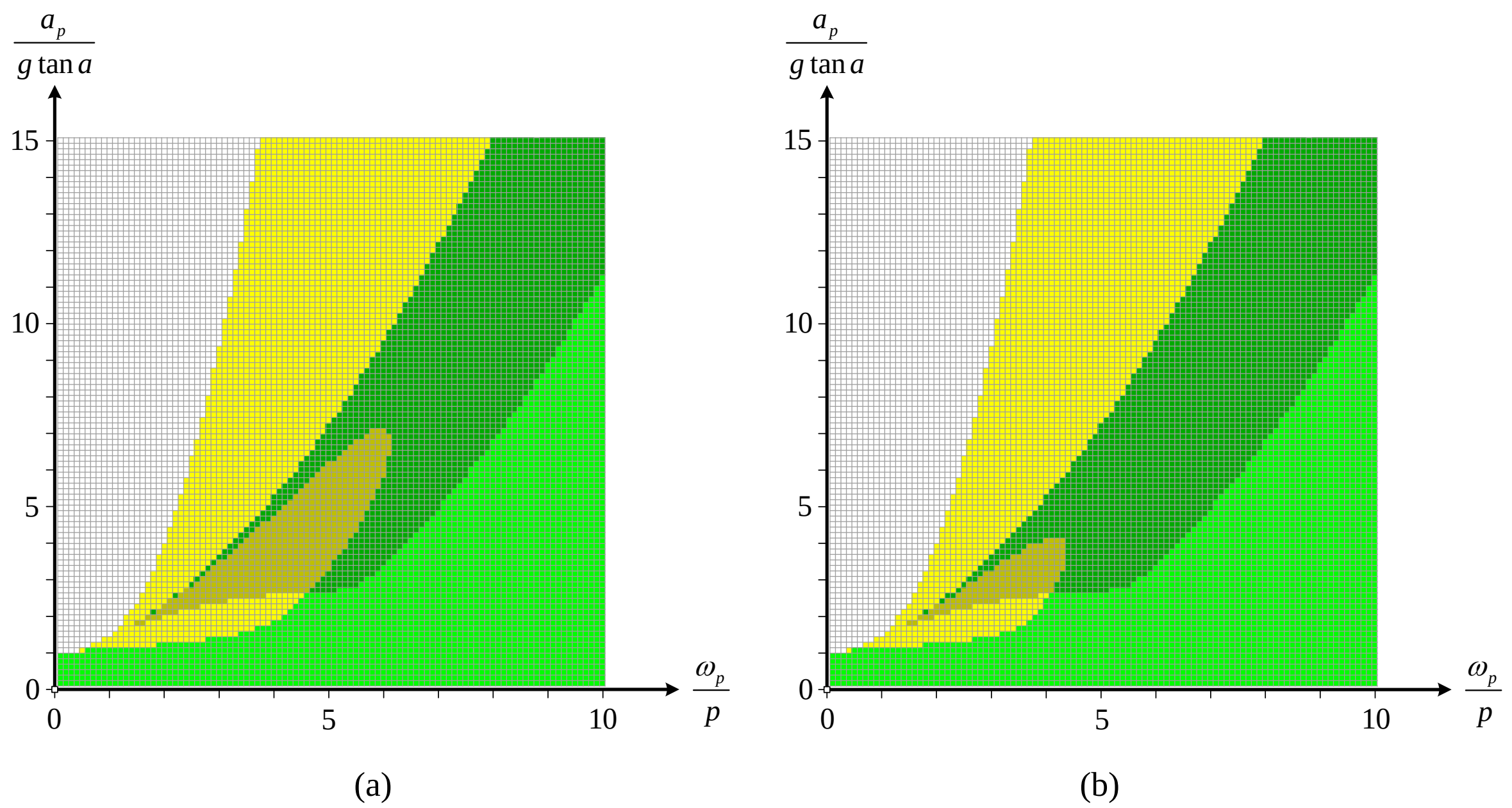

**Figure 9** Overturning acceleration spectra of a free-standing rectangular block with slenderness $a \cong 10\,deg$, frequency parameter $p \cong 1.13\,rad/s$, $r_{\max} \cong 0.912$, subjected to one-sine pulse, (a) $r = 0.85\,r_{\max}$ and (b) $r = 0.7\,r_{\max}$.

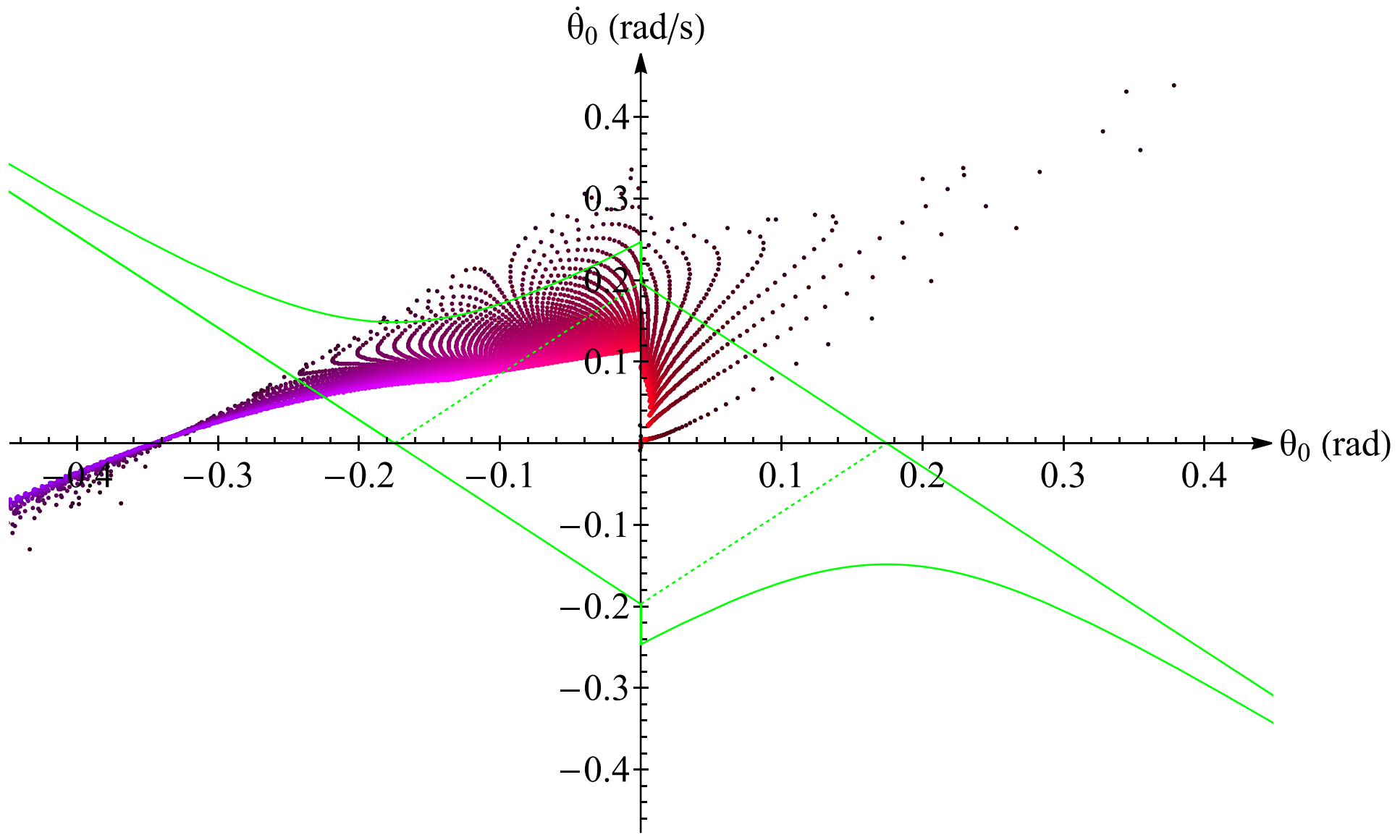

**Figure 10** Output $(\theta_0, \dot{\theta}_0)$ of the forced rocking stage for the overturning spectrum of Figure 9b (one-sine pulse) superimposed over the safe libration boundaries (rectangular block with $a \cong 1\,deg$, $p \cong 1.13\,rad/s$, $r = 0.7\,r_{\max}$).

## 4.2 Type-B pulse (one-cosine)

The acceleration, velocity, and displacement histories of a one-cosine (type-B) pulse are presented in [16]. The pulse acceleration is shown in Figure 6b and it is given by:

$$\ddot{u}_g(t) = a_p \cos(\omega_p t), \quad (\text{with } a_p > g\ \tan a), \tag{51}$$

with the duration of the forced rocking motion being equal to the period of the pulse:

$$T_f = T_p = \frac{2\pi}{\omega_p}. \tag{52}$$

Figure 11 shows the overturning spectra in a 100 × 100 pixel graph for two cases of $r$, i.e., $r = r_{\max} \cong 0.912$ and $r = 0.7r_{\max}$. The corresponding outputs $(\theta_0, \dot{\theta}_0)$ at the end of the forced rocking stage are shown in Figure 12. Regarding the overturn during the free vibration regime, each spectrum presents two distinct unsafe regions of interest, namely regions I and II, in which the block overturns without impact in the free rocking regime. However, points in region II have experienced an impact during the forced rocking regime, and thus the block enters the free vibration regime with $\theta_0 > 0$. Since now $\theta_0\dot{\theta}_0 < 0$, very few points fall within the safe regions (blades $\boldsymbol{S}$+ and $\boldsymbol{S}$-) with the rest falling in regions I and II. This shows that the influence of $r$ is rather small, i.e., reducing $r$ alters region II while leaves region I practically unaffected. This is manifested clearly in Figure 12, where it is observed that reducing $r$ modifies the output only for the points with $\theta_0 > 0$.

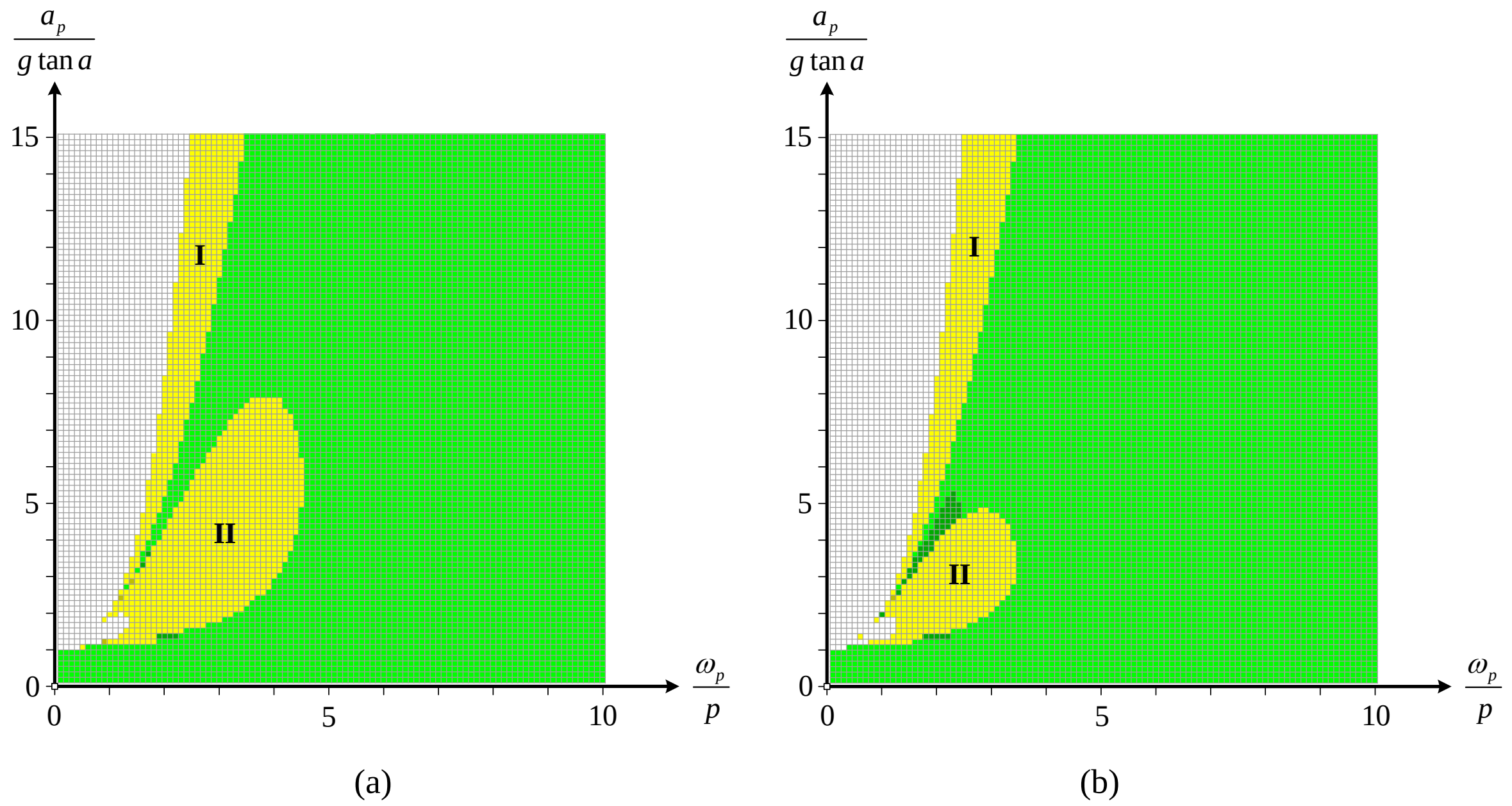

**Figure 11** Overturning acceleration spectrum of a free-standing rectangular block subjected to one-cosine pulse with slenderness $a \cong 10\ deg$, frequency parameter $p \cong 1.13\ rad/s$ and (a) $r = r_{\max} \cong 0.912$, (b) $r = 0.7r_{\max}$.

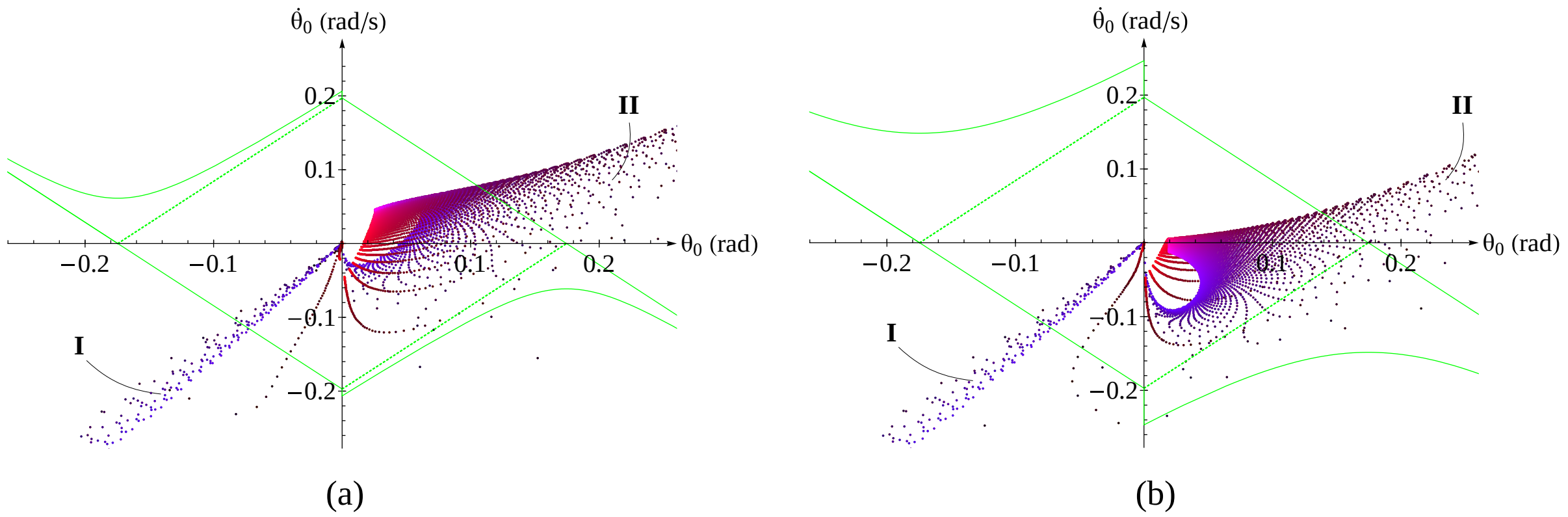

**Figure 12** Output $(\theta_0, \dot{\theta}_0)$ of the forced rocking stage for the overturning spectrum of Figure 11 superimposed over the safe libration boundaries (one-cosine pulse with $a \cong 10\ deg, p \ \cong 1.13\ rad/s$, and (a) $r = r_{\max} \cong 0.912$, (b) $r = 0.7 r_{\max}$).

## 4.3 Symmetric and antisymmetric Ricker wavelets

The scaled second derivative of the Gaussian distribution, $e^{-\frac{t^2}{2}}$, known as the symmetric Ricker Wavelet [40,41] and widely referred to as the Mexican Hat wavelet [42], will also be considered (Figure 13a). The ground acceleration is given by:

$$\ddot{u}_g(t) = a_p \left(1 - \frac{2\pi^2 t^2}{T_p^2}\right) e^{-\frac{1}{2}\left(\frac{2\pi^2 t^2}{T_p^2}\right)}. \tag{53}$$

Since the ground acceleration builds up gradually, a robust root-finding procedure has been implemented to determine the first instant when $a_p > g \tan a$ and rocking is initiated.

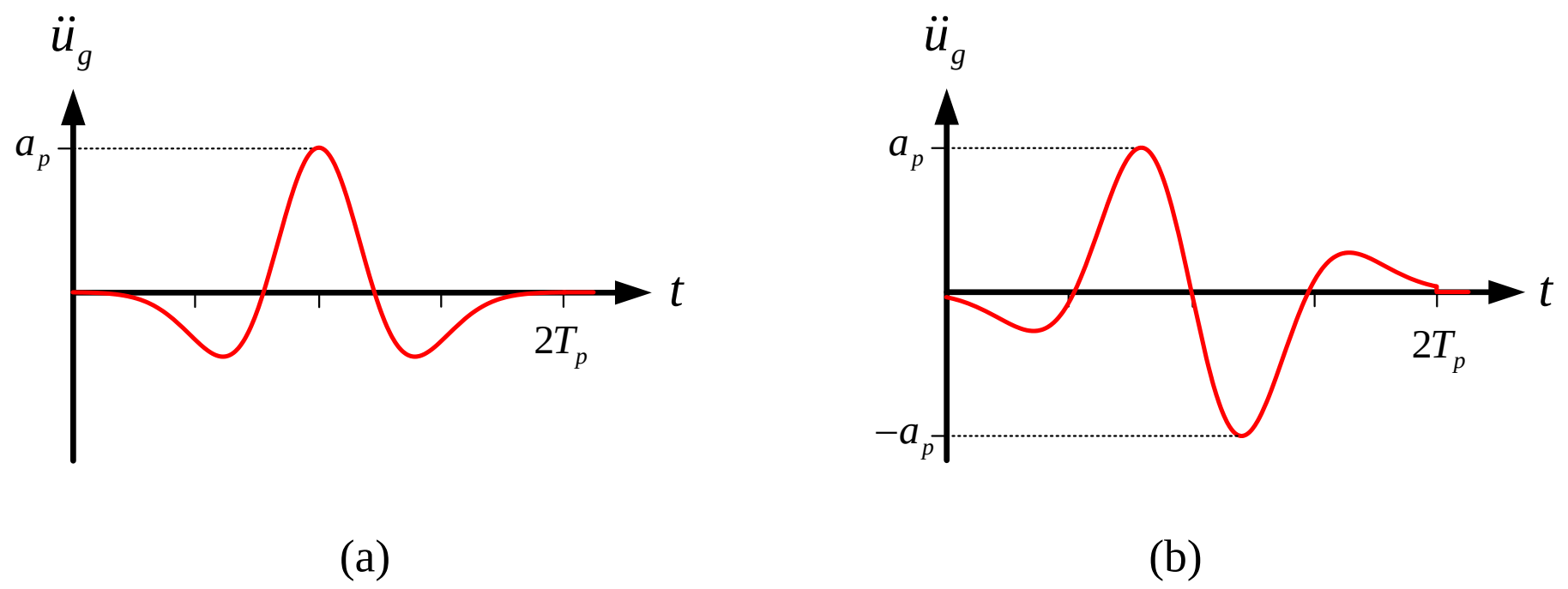

**Figure 13** (a) Symmetric Ricker wavelet (b) Antisymmetric Ricker wavelet.

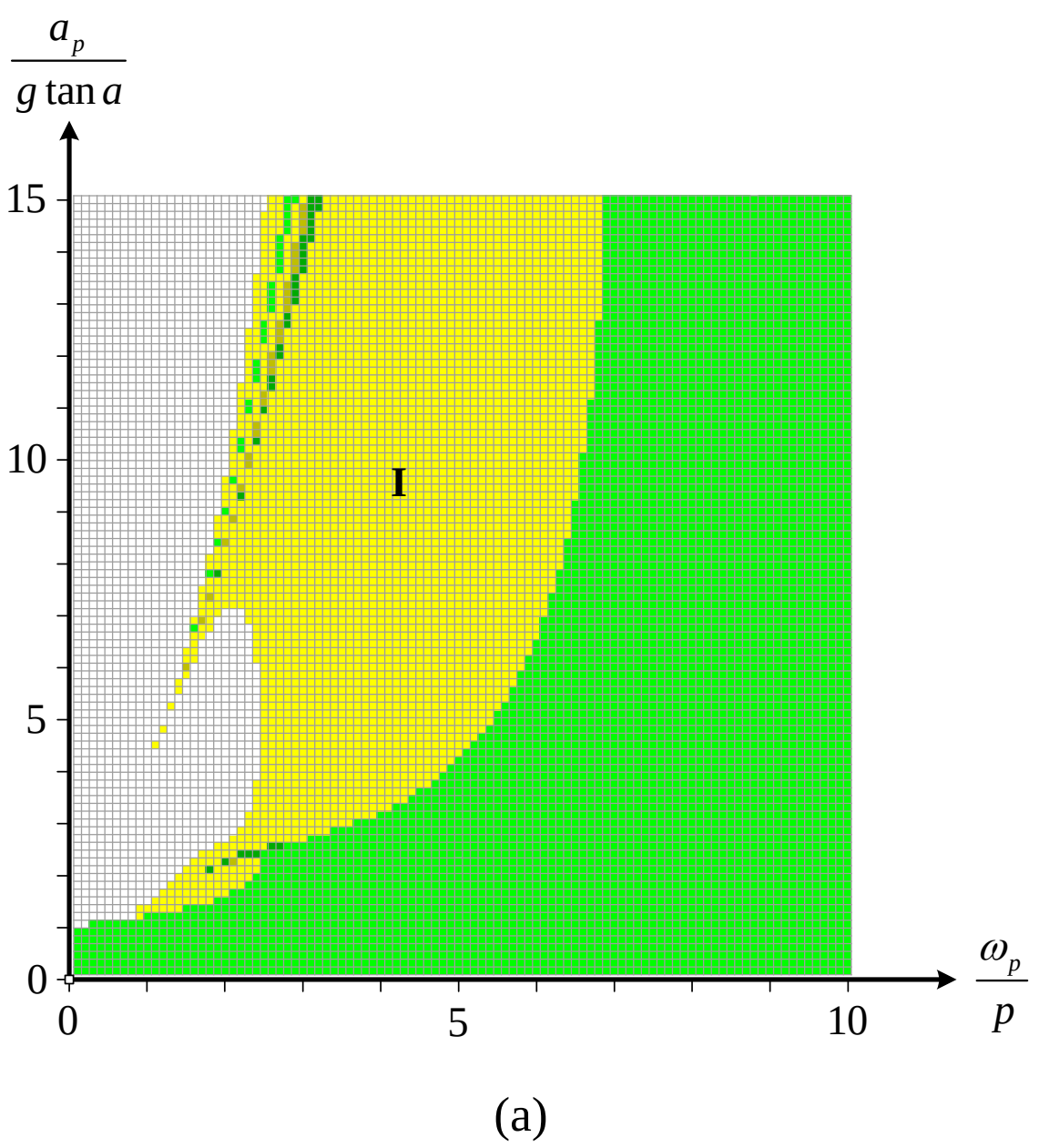

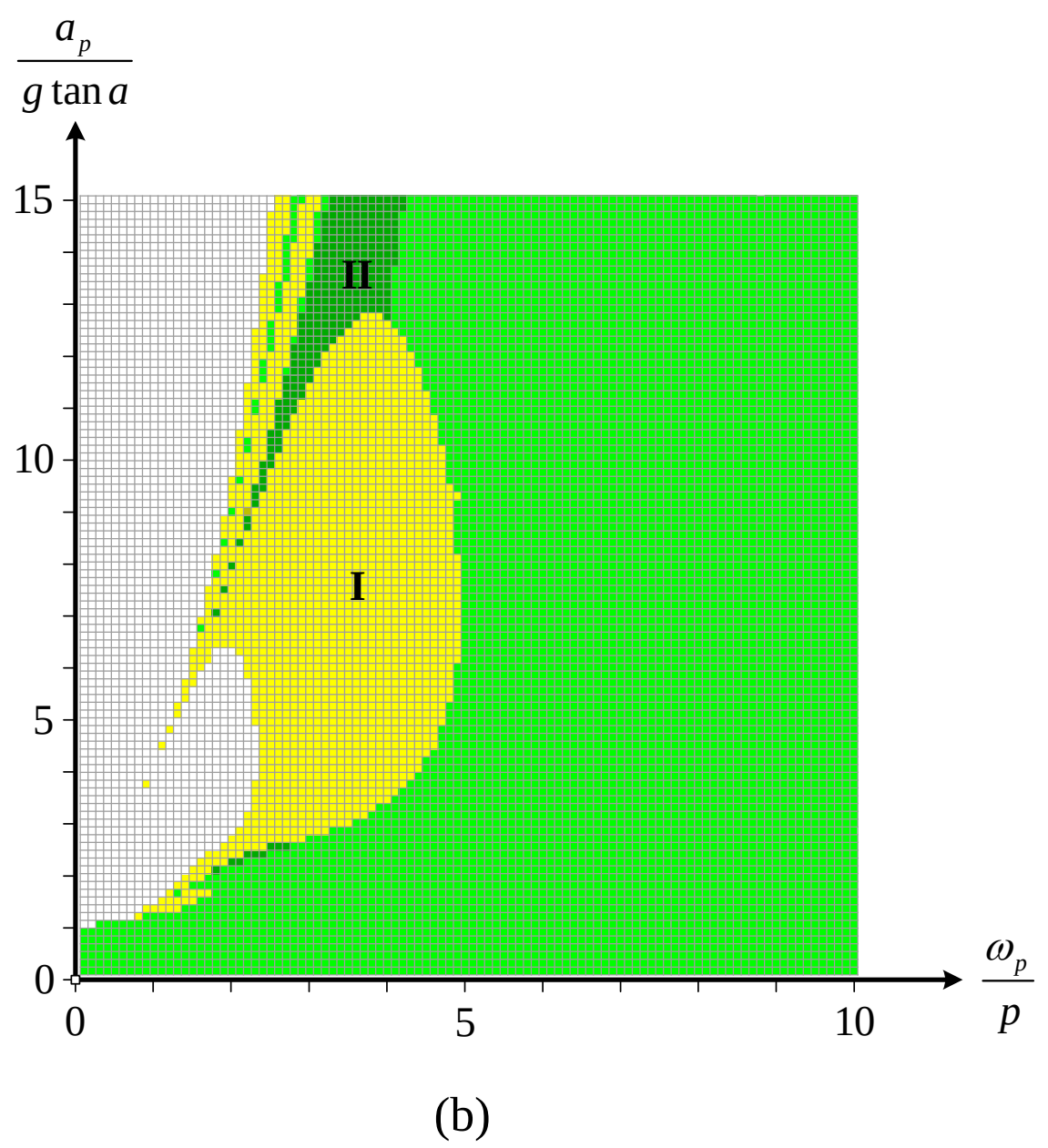

**Figure 14** Overturning acceleration spectrum of a free-standing rectangular block subjected to a symmetric Ricker pulse with slenderness $a \cong 10\, deg$, frequency parameter $p \cong 1.13\, rad/s$ and (a) $r = r_{\max} \cong 0.912$, (b) $r = 0.7r_{\max}$.

Figure 14 shows the overturning spectra in a $100 \times 100$ pixel graph for two cases of $r$, i.e., $r = r_{\max} \cong 0.912$ and $r = 0.7r_{\max}$. The overturning spectra have been constructed using the exact stability criterion (20) after the numerical solution of the forced vibration regime. The corresponding outputs $(\theta_0, \dot{\theta}_0)$ of the forced rocking stage are shown in Figure 15. The unsafe region I which is adjacent to the safe regions of both spectra corresponds to cases with $\theta_0 < 0$ and $\dot{\theta}_0 < 0$ which have experienced a single impact in the forced rocking stage; those points have landed into the ***U0***- region and therefore overturn without impact in the free rocking stage. Due to the impact, which generally occurs shortly after the peak of the ground acceleration and before the acceleration changes sign again (Figure 13a), the output $(\theta_0, \dot{\theta}_0)$ of the forced rocking stage is depended on the coefficient of restitution. The overturning spectrum of Figure 14b is different to the that of Figure 14a for two reasons; first, as mentioned, the output of the forced rocking stage is different and, second, the blade ***S+*** safe region is now considerably wider and thus encompasses more points, i.e., the safe region II of Figure 14b, which corresponds to the blade ***S+*** region of Figure 15b.

Similarly, the scaled third derivative of the Gaussian distribution, $e^{-\frac{t^2}{2}}$, known as the antisymmetric Ricker Wavelet has also been considered (Figure 13b). The ground acceleration is given by:

$$\ddot{u}_g(t) = \frac{a_p}{\beta}\left(\frac{4\pi^2 t^2}{3T_p^2} - 3\right)\frac{2\pi t}{\sqrt{3}T_p} e^{-\frac{1}{2}\left(\frac{4\pi^2 t^2}{3T_p^2}\right)}, \tag{54}$$

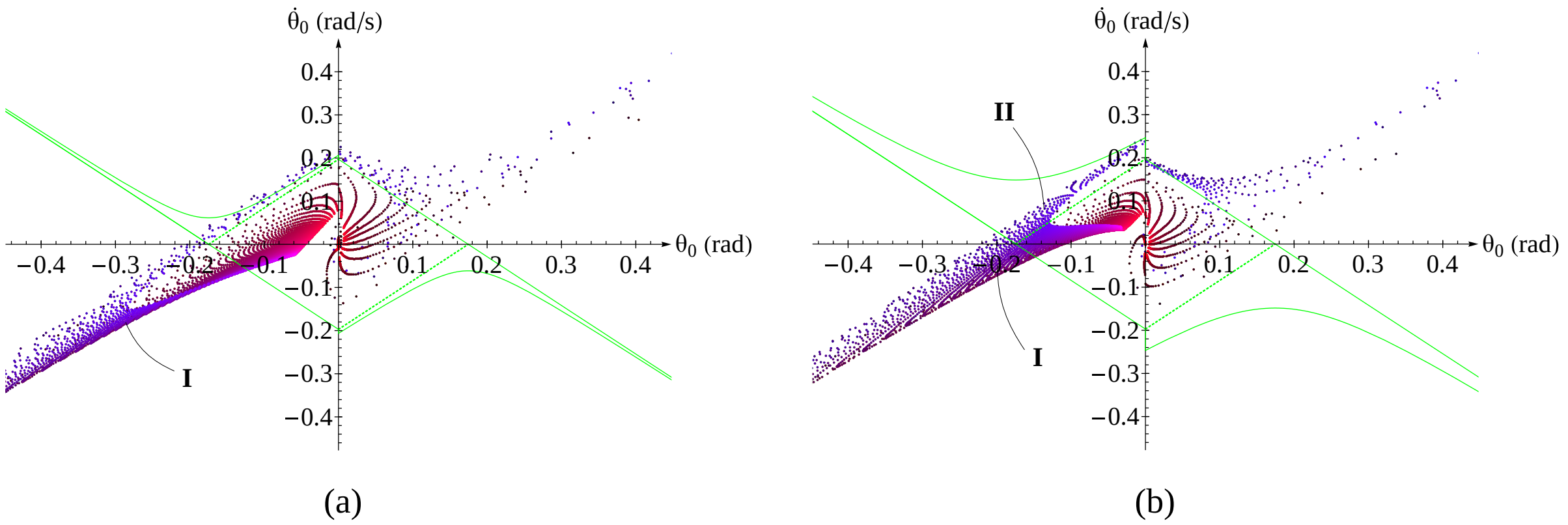

**Figure 15** Output $(\theta_0, \dot{\theta}_0)$ of the forced rocking stage (symmetric Ricker pulse) for the overturning spectrum of Figure 14 superimposed over the safe libration boundaries (block with $a \cong 10\ deg$, $p\ \cong 1.13\ rad/s$, and (a) $r = r_{\max} \cong 0.912$, (b) $r = 0.7r_{\max}$).

with $\beta$ being a factor equal to 1.38 which enforces the above function to have a maximum of $a_p$ [22]. As in the case of the symmetric Ricker wavelet, a robust root-finding procedure is implemented to determine the first instant when $a_p > g \tan a$ and rocking is initiated. The overturning spectra are shown in Figure 16, while the corresponding outputs $(\theta_0, \dot{\theta}_0)$ of the forced rocking stage are shown in Figure 17. Two unsafe regions, namely regions I and II, are identified in the spectra. The former contains points experiencing a single impact in the forced rocking stage while the latter contains points experiencing two impacts. Both of them, however, lead to overturning without impact in the free rocking stage, as they correspond to the ***U0-*** and ***U0+*** unsafe regions of Figure 4, respectively.

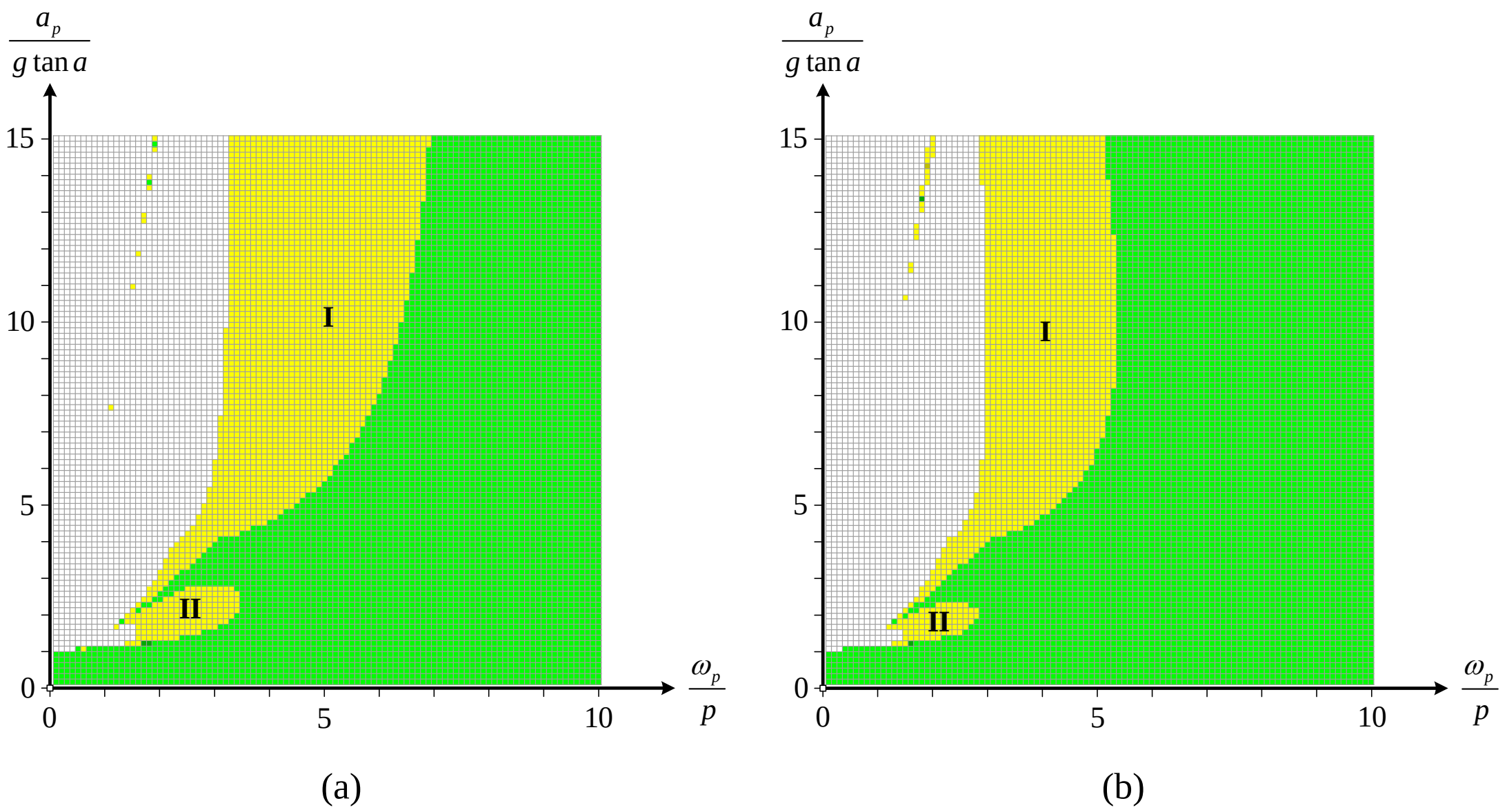

**Figure 16** Overturning acceleration spectrum of a free-standing rectangular block subjected to an antisymmetric Ricker pulse with slenderness $a \cong 10\ deg$, frequency parameter $p\ \cong 1.13\ rad/s$ and (a) $r = r_{\max} \cong 0.912$, (b) $r = 0.7r_{\max}$.

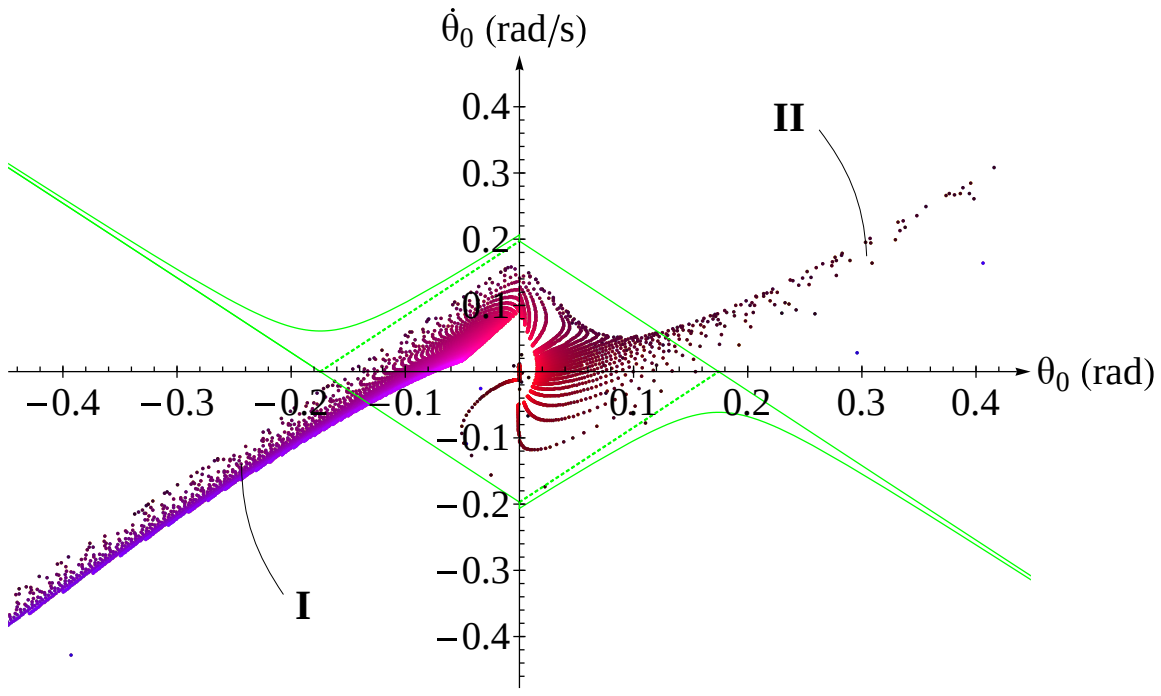

(a)

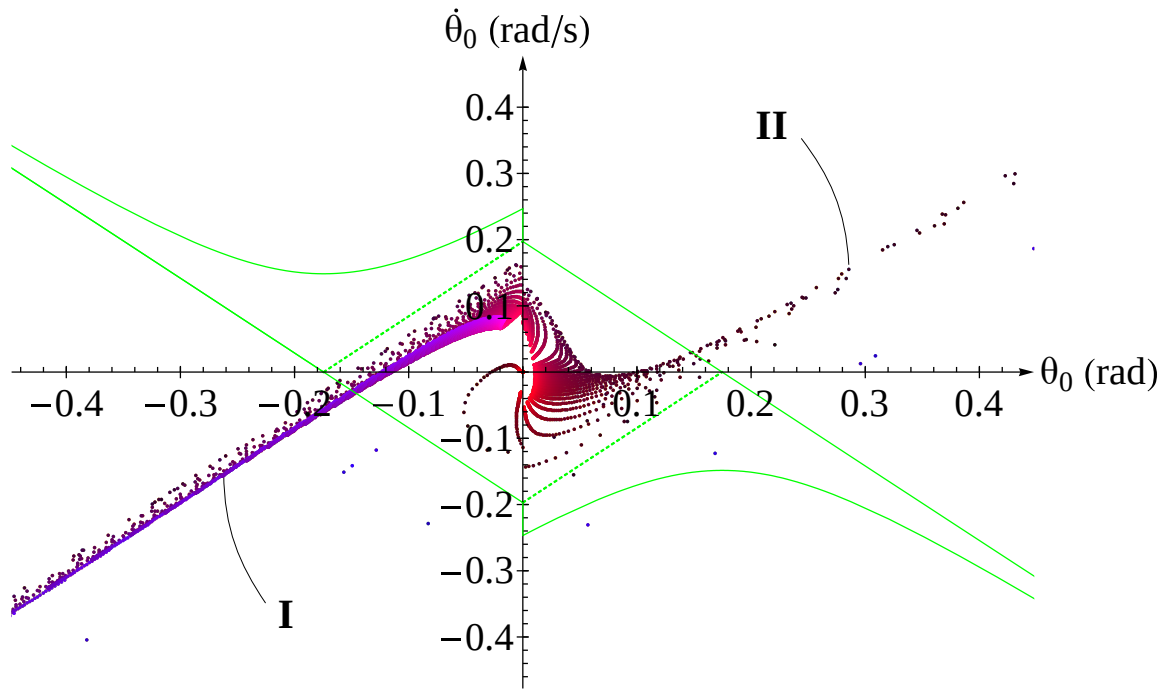

(b)

**Figure 17** Output $(\theta_0, \dot{\theta}_0)$ of the forced rocking stage (antisymmetric Ricker pulse) for the overturning spectrum of Figure 16 superimposed over the safe libration boundaries (block with $a \cong 10\ deg$, $p\ \cong 1.13\ rad/s$ and (a) $r = r_{\max} \cong 0.912$, (b) $r = 0.7 r_{\max}$).

## 5. Conclusions

This study provides the missing stepping stone in the understanding of the stability of rocking structures. The underlying energy flow in rocking dynamics was utilized to elucidate rigid blocks response details and overturning. Based solely on the law of the conservation of energy the exact criteria for a rigid block to overturn in the free vibration regime were established for the first time in the context of both nonlinear and linear theory. First, the rocking motion of rigid blocks without loss of energy during impacts was studied, resulting in a separatrix with a rhomboidal shape that corresponds to stable libration inside it while the region outside it represents the motion's toppling paths. On the other hand, the energy loss during impacts led to a novel extension of the libration zone, beyond the initial rhomboidal region, forming two antisymmetric blade-like regions admitting stable (= non-toppling) oscillations. That is, if the initial conditions of the free oscillation regime fall within the libration region, then it is *a priori* determined that the block will not overturn and *vice versa*. Subsequently, the analytical solution to the nonlinear free rocking problem is derived, utilizing integration techniques stemming from the solution of the nonlinear pendulum. The solution including energy loss during impacts is given in segments before and after impacts.

In addition, pulse-type overturning spectra for the one-sine acceleration pulse (type-A=forward pulse), the one-cosine pulse (type-B=forward-and-back pulse), and the symmetric and antisymmetric Ricker wavelets were constructed by numerical integration of the equation of motion only during the pulse duration. If the block does not topple during the forced regime, the angle and angular velocity at the end of the excitations are evaluated at the end of the pulse and the numerical integration is terminated. At this point, the safety or unsafety of the block is assessed immediately, without any further numerical integration, using the exact overturning nonlinear criterion.

## Authors' Contributions

All authors conceived the study and AC carried out the initial investigation of the problem; GT carried out the literature review; AC and GT carried out software development, as well as the validation and visualization of the results; they also contributed to writing the original draft and writing and reviewing the final version of the manuscript. PT contributed to writing and reviewing the final version of the manuscript. All authors gave final approval for publication and agree to be held accountable for the work performed therein.

## Acknowledgments

The authors would like to thank Assistant Professor Yannis Kominis for his invaluable discussions.